\documentclass[journal]{IEEEtran}
\usepackage{cite}
\usepackage{amsmath,amsfonts}
\usepackage{algorithm}
\usepackage{algpseudocode}
\usepackage{array}
\usepackage[caption=false,font=normalsize,labelfont=sf,textfont=sf]{subfig}
\usepackage{textcomp}
\usepackage{stfloats}
\usepackage{url}

\usepackage[breaklinks]{hyperref}
\usepackage{verbatim}
\usepackage{graphicx}
\usepackage{tcolorbox}
\usepackage{enumitem}
\newcommand{\orcid}[1]{\href{https://orcid.org/#1}{\includesvg[width=10pt]{orcid}}}
\usepackage{balance}
\usepackage{animate}

\usepackage{listings}
\usepackage{xcolor} 

\lstdefinelanguage{json}{
    basicstyle=\ttfamily\small,        
    showstringspaces=false,             
    breaklines=true,                    
    frame=single,                        
    backgroundcolor=\color{gray!5},     
    keywordstyle=\bfseries\color{blue}, 
    stringstyle=\color{teal},           
    numberstyle=\color{purple},         
    commentstyle=\color{gray},          
    morestring=[b]",                     
    morecomment=[l]{//},                 
    morestring=[s]{\{}{\}},              
    morestring=[s]{[}{]},                
}

\ifCLASSINFOpdf
\else
\fi
\begin{document}

\graphicspath{{figures/}}
%
\title{Agentic Search Engine for Real-Time IoT Data}
%
%
%

\author{Abdelrahman Elewah,
        and~Khalid Elgazzar
\thanks{A. Elewah and  K. Elgazzar are with the IoT Research Lab https://iotresearchlab.ca/, the Department of ECSE, Ontario Tech University, 2000 Simcoe Street North Oshawa, Ontario L1G 0C5 Canada (e-mail: {abdelrahman.elewah,khalid.elgazzar}@ontariotechu.ca).}
}

\maketitle

\begin{abstract}

The Internet of Things (IoT) has enabled diverse devices to communicate over the Internet, yet the fragmentation of IoT systems limits seamless data sharing and coordinated management. We have recently introduced SensorsConnect, a unified framework to enable seamless content and sensor data sharing in collaborative IoT systems, inspired by how the World Wide Web (WWW) enabled a shared and accessible space for information among humans. This paper presents the IoT Agentic Search Engine (IoT-ASE), a real-time search engine tailored for IoT environments. IoT-ASE leverages Large Language Models (LLMs) and Retrieval Augmented Generation (RAG) techniques to address the challenge of searching vast, real-time IoT data, enabling it to handle complex queries and deliver accurate, contextually relevant results. We implemented a use-case scenario in Toronto to demonstrate how IoT-ASE can improve service quality recommendations by leveraging real-time IoT data. Our evaluation shows that IoT-ASE achieves a 92\% accuracy in retrieving intent-based services and produces responses that are concise, relevant, and context-aware, outperforming generalized responses from systems like Gemini. These findings highlight the potential IoT-ASE to make real-time IoT data accessible and support effective, real-time decision-making.

\end{abstract}

\vspace{0.1cm}
\begin{center}
\small Code is available at: \url{https://github.com/SensorsConnect/IoT-Agentic-Search-Engine}
\end{center}
\vspace{0.1cm}


\IEEEpeerreviewmaketitle

\section{Introduction}

\IEEEPARstart{T}he Internet of Things (IoT) has reshaped human life through the growing number of connected things. Several domains have leveraged the implementation of the IoT, from home appliances and wearable devices to intelligent transportation and logistics systems. By the end of 2025, 75\% of devices \cite{explodingtopics2024iot} will be connected to the internet. IoT devices are the nodes between the physical and digital domains. Conceptually, IoT devices \cite{elgazzar2022revisiting} act as data sinks receiving data to control connected actuators or data sources sending data collected from embedded sensors. Therefore, they have become the primary real-time data sources/drains that represent a paradigm shift in real-time decision-making systems. Based on the statistics \cite{explodingtopics2024iot}, the data generated or consumed by IoT devices will exponentially grow in the coming years. As a result, the demand for a real-time search engine for the collected IoT data will increase to keep pace with the needs of real-time decision-making.

Multiple attempts \cite{IoTSearch1, IoTSearch2} in the literature endeavored to provide a search engine for IoT data. Shodan \cite{shodan,  mulero2023detection}, a search engine for devices connected to the internet, crawls and indexes publicly available metadata of internet-connected devices. Shodan indexes data of a connected device, such as the type of the device, software used, the name/type of the deployed server and its version using devices' banners. Thus, Shodan serves as a search engine to collect statistics \cite{liang2019search} on connected devices in a specific region or gain insight into which software products are trending for market research purposes. Like Shodan, Censys \cite{censys} and Reposify\_ acquired by CrowdStrik \cite{crowdstrike2024} use crawling over the metadata of the connected devices to build their search engines. In contrast, these search engines focus on the security aspect of the connected devices. Organizations and enterprises can use these search engines to gain insight, such as determining vulnerable devices, assessing potential risks, and monitoring and scanning assets connected to the Internet. Thus, Shodan, Censys, or Reposify can retrieve only publicly available metadata of IoT devices, and we consider these types of search engines as device discovery tools. Based on the adopted approach (crawling over devices' metadata), if these search engines try to extend their service to leverage the application data driven by connected devices, they will not access it due to the implemented authentication protocol restricting unauthorized connectors. Furthermore, assuming they gain the authority to access the application data, the fragmentation of the IoT devices and the heterogeneity of their data model pose further challenges.

IoTCrawler \cite{iggena2021iotcrawler} is a framework designed to enable searching for the data produced by IoT applications. IoTCrawler uses algorithms adopted in web search engines like crawling, indexing, ranking, and discovering. Integrating IoT-Stream \cite{elsaleh2020iot}, which serves as a semantic annotation layer, IoTCrawler's scope is enabling the search engine for the existing IoT systems and addressing the challenge of the heterogeneity/fragmentation of IoT systems.

The IoTCrawler group has made diligent efforts to enable searching for IoT data; however, IoTCrawler has not been widely adopted by the IoT community. Assuming IoTCrawler has created adaptors for all existing IoT protocols, such as Message Queuing Telemetry Transport (MQTT), Constrained Application Protocol (CoAP), Hypertext Transfer Protocol/Secure (HTTP/HTTPS), and Advanced Message Queuing Protocol(AMQP) etc, the data/ payloads used over these protocols have no consensus between IoT systems. Thus, integrating the sensing data of a specific IoT system in IoTCrawler requires building its dedicated interface. In other words, enabling the sensing data of IoT systems through IoTCrawler requires handling endless Application Programming Interface (APIs), which is considered infeasible.

We have recently introduced SensorsConnect \cite{SensorsConnect}, which mimics the World Wide Web (WWW) but for IoT devices. Thus, using a unified transfer protocol and messaging standard like HTTP and HTML used in WWW is applicable. Additionally, two case studies \cite{SensorsConnect} engaging drive-through coffee shops and the US-Canada border demonstrated that utilizing a search engine for real-time IoT data in service recommendation applications reduces the average service time by 46\% and 31\%, respectively.

The growing number of websites on the WWW \cite{brin1998anatomy} poses a searching challenge, and since the web content follows unstructured patterns, search engines \cite{brin1998anatomy} use the hyperlink feature for crawling, indexing, and ranking the WWW content. Similarly, 41.6 billion is the expected number of  IoT devices that will generate 80 zettabytes of data by 2025; hence, SensorsConnect will face the search challenge. We designed SensorsConect to support searching natively and avoid requiring deploying crawling indexing and ranking algorithms. Instead, SensorsConnect supports using cutting-edge search approaches that recently emerged in the WWW and may replace the traditional search approaches \cite{spatharioti2023comparing}. Shedding light on the search engine component of SensorsConnect, the paper introduces a real-time IoT Agentic Search Engine (IoT-ASE) that leverages the Large Language Models (LLM) \cite{minaee2024large} models and Retrieval Augmented Generation (RAG) \cite{lewis2020retrieval} Approaches. 

The remainder of the paper is organized as follows. Section \ref{sec:Background&Related work} briefly discusses the related works. Section \ref{sec:Architecture} delineates the architecture of IoT- Retrieval-Augmented Generation- Search Engine (IoT-RAG-SE) and a Generic Agentic RAG (GA-RAG) workflow for the agentic RAG systems. It also presents the implementation of IoT-ASE based on GA-RAG. Section \ref{sec:Use Case} elaborates on evaluating IoT-ASE using a real-life case scenario and discusses some experimental results. Lastly, Section \ref{sec:Conclusion} provides concluding remarks.

\section{Background and Related Work}
\label{sec:Background&Related work}

\subsection{Background}
\label{sec:Background}

Looking deeply into the root challenge hindering the presence of IoT search engines supporting IoT application data, we found that the underlying causes are the fragmentation of IoT systems and the heterogeneity of sensing data. By nature,  IoT systems have taken intranet shapes for security and privacy considerations, and external integrations require a particular API for each IoT system. Historically, web search engines were invented as solutions to sort and index the contents of the World Wide Web (WWW) framework \cite{berners1994world}, which was initially introduced as a collaborative framework to share documents for humans. No doubt, many IoT application requires certain levels of privacy and security. Nonetheless, we overlooked the usefulness of having a collaborative framework for publicly available IoT data, especially if we relate how WWW has changed human life. SensorsConnect \cite{SensorsConnect} provides a unified infrastructure for IoT systems so we can adopt standards and protocols for sharing IoT sensing data to pave the way for building an IoT search engine. The SensorsConnect framework has been designed to support querying IoT data by the end users and the IoT systems integrated into the framework. 

\subsubsection{SensorsConnect Framework}
\begin{figure}
\centerline{\includegraphics[width=0.5\textwidth]{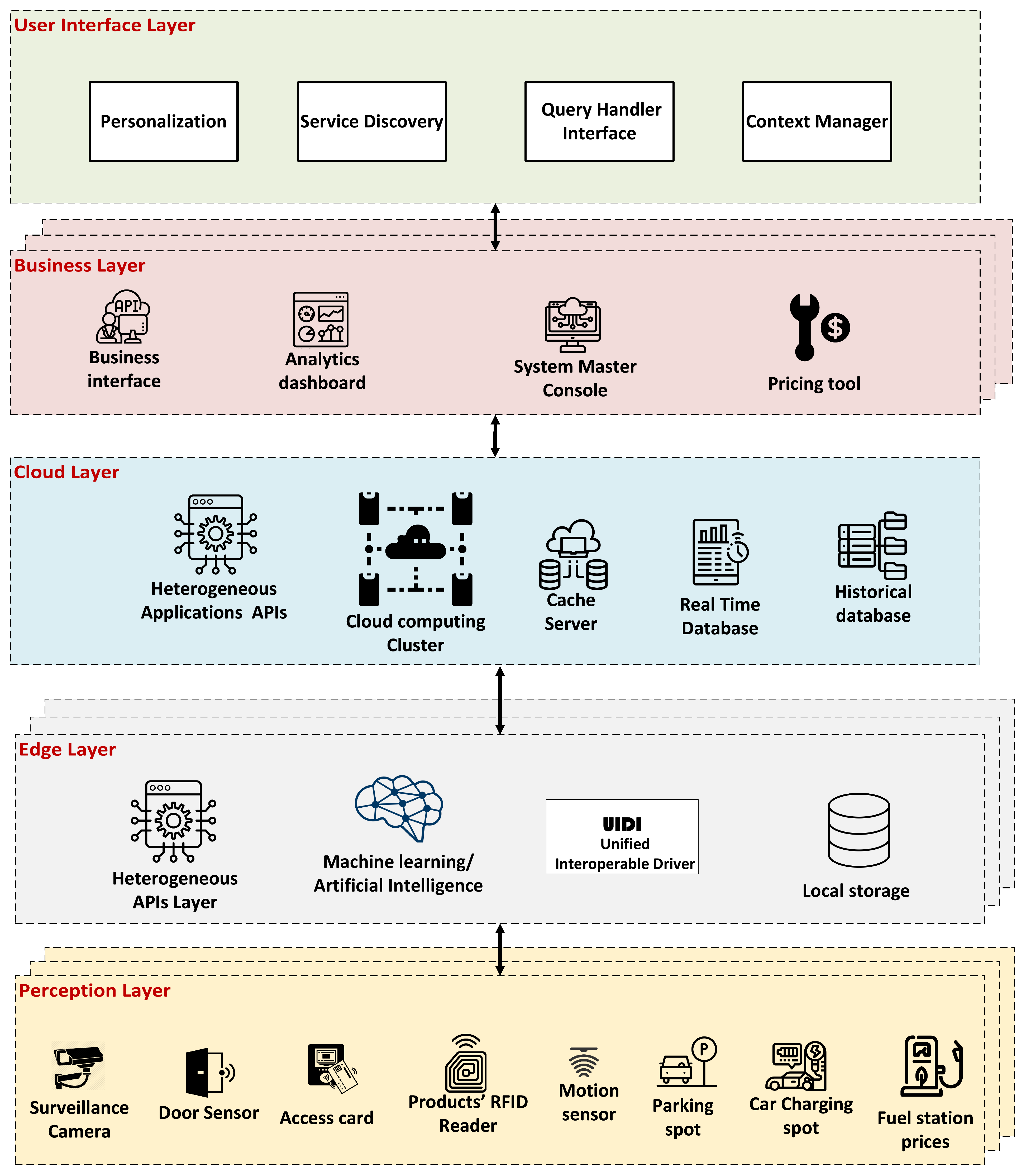}}
\caption{SensorsConnect Architecture}
\label{fig:ps-arch}
\end{figure}

\hfill\\ 
Fig. \ref{fig:ps-arch} depicts the SensorsConnect architecture, combining its five layers: perception, edge, cloud, business, and user interface.
The perception layer interfaces with the physical domain and handles the sensing devices. The edge computing layer is an adapter that formats heterogeneous sensing data in the Unified Interoperable Driver for IoT (UIDI) \cite{elewah2022thingsdriver} standard that mimics the HTML standard in WWW during its early stages. Thus, sharing content/IoT data between integrated IoT systems is applicable without requiring a dedicated driver for each connection. In certain cases, machine learning (ML) and artificial intelligence (AI) models hosted locally on the edge layer carry out data extraction tasks. For instance, relying on these components, the edge layer can \cite{Traffic} infer the traffic volume from a surveillance camera at an intersection and send the extracted data in UIDI format rather than sending the entire video stream to the cloud layer.

The Cloud layer manages IoT data of the integrated services in SensorsConnect using three components: 1) The real-time database stores the last update received from IoT devices. 2) The historical database stacks the time series of the collected data. 3) The cache server keeps retrieved data, which is most likely recalled again shortly, so this can speed up the retrieval time and save resource usage. Additionally, SensorsConnect can host multiple cloud functions for a single IoT system. Hence, this set of cloud functions enables IoT systems to manage received data from their IoT devices, interpret and add metadata for the perceived data, especially from limited-constrained devices, supervise collaboration of IoT systems and share IoT sensing data among them, and finally, serve queries of the end users. 

The business layer provides tools for enterprises and organizations to manage their IoT systems integrated into the framework. These tools include: 1) a master management console to access the devices installed in any layer, 2) an analytic dashboard to monitor the states of devices and detect vulnerabilities, if any, 3) a business interface to enable integration of external services like completing transactions of trade operations or booking parking spots., and 4) finally, Pricing tool suggest convenient pricing scheme based on business need. 

Lastly, the user interface layer contains a query handler, service discovery \cite{ServicesDiscovery},  context manager and personalization to get user queries based on user preferences \cite{elgazzar2014daas}, discover new services that match user needs, extract context information, and personalize the user experience.

SensorsConnect maintains primarily two workflows: Collect-Store and Query-Respond. On the one hand, the Collect-Store cycle process manages data collection in the perception layer, data unification in the edge layer, and data storing at the data management component in the cloud layer. On the other hand, the Query-Respond cycle triggers when a user submits a query through the user interface layer, passing through the query understanding component deployed in the cloud layer and retrieving IoT data matching user intent from the data management component. This paper focuses on realizing the Query-Respond workflow, and since the data management component overlaps between the two workflows, the paper also presents a unified data model that handles the heterogeneity of IoT data and meets the needs of both workflows.

\subsection{Related Work}
The World Wide Web faced challenges in searching and organizing content in its infancy stages. Crawling, indexing, and ranking approaches \cite{brin1998anatomy} have been used to overcome searching for content in an unorganized world of content. Researchers in the natural language processing (NLP) field have recently presented methodologies to tackle searching tasks with techniques that achieve remarkable performance. The architecture of SensorsConnect \cite{SensorsConnect} facilitates integrating search engines that leverage these approaches and overcoming limitations in traditional search methods.

\subsubsection{Large Language Model}\hfill\\ 
 Recently, there has been a significant breakthrough in the Natural Language Processing field grounded in the evolution of the large language models (LLM) \cite{minaee2024large} based on transformers \cite{vaswani2017attention} and trained on Web-scale content. For instance, the capabilities of rule-based chatbots, like Google Assistant, Siri, and Alexa, are limited to predefined user intents. Once a user triggers one of these intents, the rule-based chatbot responds with predefined answers containing the data the user is looking for. With the LLM breakthrough, the ChatGPT application based on GPT-3.5, or GPT-4 \cite{achiam2023gpt} models have unprecedented abilities, including (1) generating like-human answers that are not explicitly predefined in the model, (2) solving general tasks such as generating programming codes, (3) following user guidelines for complicated novel tasks, (4) performing multi-step logical solutions if needed, especially in complex tasks, (5) Learning new tasks using group of examples provided in the user prompt. Therefore, LLM models become the cornerstone of general-purpose AI agents.

However, the LLM model like GPT-4  can't be a stand-alone solution for the search engine component of SensorsConnect due to the following limitations:
\begin{enumerate}
    
\item  Since LLM models are trained on mixed information, true and false facts, they tend to hallucinate. These models can generate well-convincible answers but are totally incorrect.
\item LLMs are memoryless. For instance, the LLM models in a conversation with a user cannot memorize what the user says in a previous prompt.
\item LLM models can efficiently answer general questions; however, they struggle to provide answers for uncommon information. In other words, the performance of LLM models degrades when retrieving long-tail knowledge in a specific field.
\item  LLM models are trained in historical data and can't access external data sources, such as sensing devices/ IoT devices in SensorsConnect. LLM models are not aware of the time when a user sends a prompt and have no entry to any data that wasn't used in the training process.
\end{enumerate}

\subsubsection{Retrieval-Augmented Generation RAG}\hfill\\ 
LLMs achieved state-of-the-art performance compared with other NLP tasks due to their ability to store knowledge and information within their parameters. However, as mentioned, LLMs hallucinate, can't memorize uncommon knowledge (long-tail knowledge), and don't have access to external data sources, including real-time data that SensorsConnect hosts. Lewis et al. \cite{lewis2020retrieval} tackled many LLM limitations by proposing the Retrieval-Augmented Generation (RAG) framework. RAG augments an LLM model with a retrieval component that privileges LLM to access an external database of documents and retrieve related information during inference time to enlighten and boost the generative process. RAG combines two parts: 1) The retriever that finds relevant chunks of documents in the augmented external database based on the similarity between retrieved documents and the user input, and 2) the generator that uses the user input alongside the retrieved documents from the external database to generate an improved response. Thus, this RAG approach \cite{li2022survey} dramatically improves the LLM's capability to generate coherent, contextually reasonable, accurate, and knowledge-grounded answers. 

RAG techniques have recently been broadly adopted in commercial applications called co-pilots or LLM agents, which substantially rely on RAG stacks to provide their services. This survey \cite{zhao2024retrieval} classified RAG applications into eight classes: Text, coding, Knowledge,  Image,  Video, Audio, 3D and Science. Nonetheless, these categories do not present a solution for search engines to access real-time sensing data demanded by SensorsConnect. Moreover, the existing RAG stacks embedded the external data source once, and in the scenario of IoT-ASE, the embedded contents need to be live-updated, which requires heavy computing resources. Thus, the introduced IoT-ASE uses an IoT RAG stack that manages real-time sensing data without embedding.

\subsubsection{Agentic LLM System}\hfill\\ 
 Multi-agent systems, also known as Agentic AI systems, have recently emerged as a powerful approach for boosting the ability of LLM models to solve complex tasks. Instead of having a single LLM agent carrying on the required task, multi-agents collaborate to improve the efficiency of handling the tasks and reduce the hallucinations of LLMs.  ReAct \cite{yao2022react}, sparks the presence of agentic LLM systems, is an approach that iteratively braids reasoning and acting. The reasoning agent generates, traces and edits action plans, whereas the acting agent executes and interfaces with external sources, such as knowledge graph systems or environments, to collect more information. Other agentic systems have been introduced for specific domains such as economy, social sciences and psychology, health, and more. For instance, CollabStory \cite{venkatraman2024collabstory} is an agentic system that forms a collaborative flow to generate stories. Also, Tradinggpt \cite{li2023tradinggpt} engages in conversations and debates to act as a core of decision-making in stock and fund trading based on the insights driven from the discussions. This topic still needs further investigation, as we have no clear approach to determining the optimal design for different use cases. Thus, in this paper, we present a Generic Agentic RAG (GA-RAG) workflow.


\subsubsection{Sensors Data Retrieval Based on LLM}\hfill\\ 
The framework proposed in ref \cite{berenguer2024leveraging} leverages Large Language Models (LLMs) for building a system for dataset retrieval. The framework \cite{berenguer2024leveraging} has two key components: 1)  A deployed LLM extracts sensor data in a structured format like CSV from the sensor data presented in the HTML format on online web pages. 2) another incorporated LLM generates word embedding representations of the extracted tabular data to facilitate semantic searching. The author evaluated a list of LLM in terms of precision at one (P@1) and Mean Reciprocal Rank (MRR) and concluded that the base-large-en-v1.5 model scored at P@1: 0.9 and MRR: 0.94 outperforms the other models examined in the evaluation. 

However, this dataset retrieval framework does not handle real-time data retrieval. The experiment was conducted in historical data where 90\% of datasets were saved as the targeted data to be retrieved, and the remaining 10\% were used as input or as a query table. Furthermore, as mentioned earlier, the framework deploys an LLM model to generate a word embedding representation for the collected sensor data, which usually have numerical values, and the word embedding process generates vector (numeric) representations for words and sentences. Therefore, deploying LLM to convert or retrieve data and complement tables with the same columns is resource-intensive. Also, the author faced challenges such as finding datasets, the heterogeneity of the web pages, and API limits, so even the LLM that handles collecting sensor data from the web pages has to deal with these challenges. Therefore, we have introduced the SensorsConnect framework that 1) provides a real-time search engine for sensor data, 2) uses a unified interface for IoT devices to tackle the heterogeneity challenge, and 3) applies an approach to save the collected data so it can be smoothly stored by IoT devices and retrieved by LLM agents in real-time.

\newpage
\section{System Architecture}
\label{sec:Architecture}

\begin{figure*}
\centerline{\includegraphics[width=1\textwidth]{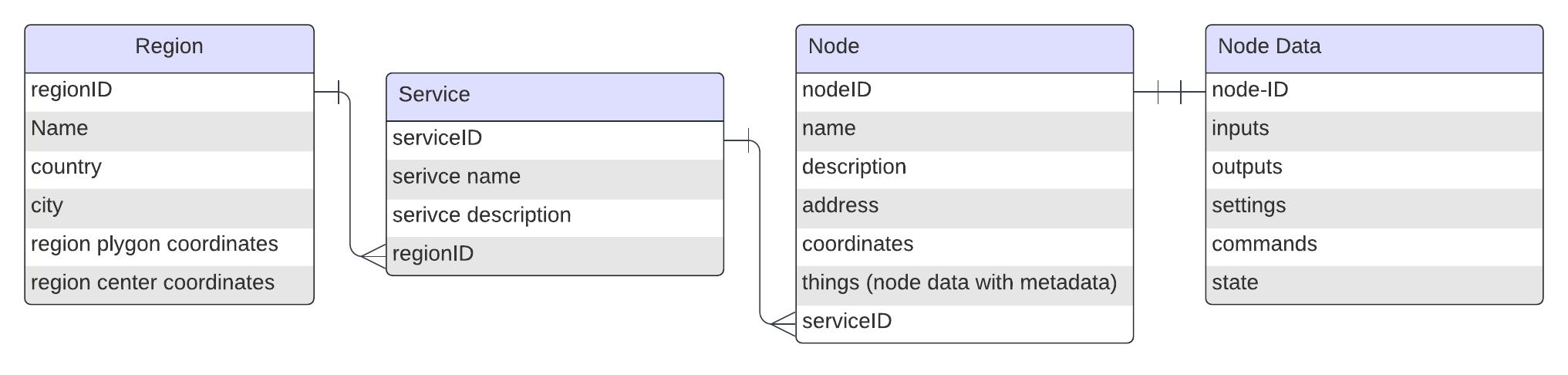}}
\vspace{-15pt}
\caption{IoT-ASE Logical Data Model}
\label{fig:data-model}
\end{figure*}

This section illustrates the following architectural components:
\begin{enumerate}
    \item The unified data model, Fig. \ref{fig:data-model}, that tackles the heterogeneity of IoT data.
    \item IoT- Retrieval Augmented Generation- Search Engine (IoT-RAG-SE), Fig. \ref{fig:RT-IoT-Engine} handling IoT query and retrieving real-time IoT data.
    \item Generic Agentic-RAG (GA-RAG), Fig. \ref{fig:state-diagram} defines a systemic approach for building RAG workflow.
    \item IoT- Agentic Search Engine (IoT-ASE), Fig. \ref{fig:Agentic-IoT} based on GA-RAG.
\end{enumerate}

\subsection{Data Model}
Humans interact with generated web content by leveraging their ability to interpret information and benefiting from advancements in User Experience (UX) and User Interface (UI) design, despite the diversity in website structures. Regardless of the used device, as long as it has a web browser, the content is rendered uniformly, relying on standardized transfer protocols (HTTP) and data representation formats (HTML, CSS, and JavaScript).

In contrast, IoT data possesses a fundamentally different nature. A vast array of sensors and connected devices continuously generate real-time data, often represented as high-dimensional time-series vectors. While IoT systems commonly rely on transfer protocols such as MQTT, CoAP, and HTTP, there is no standardized approach to defining payloads, unlike the universal adoption of HTML in the World Wide Web. This lack of standardization introduces unique challenges in handling and interpreting IoT data.

Building on this foundation, SensorsConnect simplifies IoT communications by standardizing five core topics in the UIDI messaging protocol: input, output, setting, command, and state. These topics align with key IoT device operations to enable consistent and predictable communication. The input topic receives and handles sensor data, the output topic is for actuations or responses, the setting topic handles configuration parameters, the command topic facilitates external control, and the state topic communicates the device's operational status. This streamlined approach mirrors HTML's role in the WWW, providing a universal, lightweight protocol that simplifies development and fosters seamless IoT integration across diverse applications.

We conducted an intensive quantitative analysis \cite{Hossam2024} for the existing data model in handling heterogeneous IoT data. First, we determined the requirement of IoT devices in terms of collaboration in sharing IoT data and the ability of IoT devices with limited constraints to process contents. We then investigated the requirement of LLM agents in terms of the ability to query, understand, and generate content. We found that the proposed \textit{IoT Agentic Search Engine} can adopt the document data model depicted in Fig. \ref{fig:data-model} to manage the collected data, satisfying both sides (IoT devices and LLM models). 

Given the prevalence of IoT devices in our physical environment, queries are often geographically based, and managing substantial volumes of real-time IoT data becomes critical. To address this, we propose a hierarchical IoT data model starting with the region entity for spatial partitioning, where the data model organizes services within each region, with multiple IoT devices deployed in the same area. The node data entity encapsulates abstracted IoT data, transmitted efficiently via the UIDI messaging protocol to accommodate resource-constrained devices. Meanwhile, the node entity provides descriptions and metadata, ensuring the data is interpretable by large language models (LLMs) for enhanced system intelligence and usability.

\subsection{IoT-Retrieval-Augmented Generation-Search Engine (IoT-RAG-SE)}
IoT-RAG-SE handles heterogeneous real-time IoT sensing data and routes user queries based on time and spatial factors. Utilizing the Sentence-BERT \cite{reimers2019sentence} method and a vector database based on the Hierarchical Navigable Small World (Hierarchical NSW, HNSW) \cite{malkov2018efficient} approach, as illustrated in Fig. \ref{fig:RT-IoT-Engine}, IoT-RAG-SE carries out two main processes: services descriptions embedding and performing semantic search.
\subsubsection{\textbf{Services descriptions Embedding}}
To generate high-dimensional embeddings for each service description stored in the service entity defined in the data model, Fig. \ref{fig:data-model}, it undergoes a comprehensive series of steps explained as follows:
\begin{enumerate}[label=\roman*]
    \item \textbf{Tokenization:} Firstly, a pre-trained tokenizer breaks down the service description into pre-defined tokens and generates token IDs and an attention mask for each sentence in the service description. 

    \item \textbf{Embedding:}Then, using the attention mask and token IDs, the embedding model generates dense vector representations that capture the semantic meaning of each sentence in the original service description. 

    \item \textbf{Pooling:} Following embedding, the dense vector representation is processed through a mean pooling operation, considering the attention mask provided by the tokenizer. It averages the embedding of tokens, weighted by the attention mask, to generate a single embedding vector implying the overall semantic meaning of the service description, taking into account the relevancy of tokens. 

    \item \textbf{Normalization:} Next, this single embedding vector is normalized to ensure the embeddings of the services' descriptions are consistent since it's crucial for accurately performing the cosine-similarity search in the query-semantic-search process. 

    \item \textbf{Storing in VectorDB:} Finally, the normalized embeddings vector alongside the service name is indexed in the vector database to enable fast retrieval relying on the nearest-neighbour search.
\end{enumerate}

\subsubsection{\textbf{Performing semantic search}}
The semantic search process is triggered when IoT-RAG-SE receives a query looking for a service or IoT data. First,  the query passes through the same data pipeline: tokenization, embedding, pooling, and normalization process,  generating the normalized embedding vector of the query. Then, a Hierarchical NSW, HNSW  \cite{malkov2018efficient}  semantic search is performed within the vector database between the query and the stored services descriptions to determine the top k similar services. Finally, with the top k service names and the query location, the real-time IoT database routes the query, returning nearby IoT data documents, which are interpretable by LLM models.

\begin{figure*}
\centerline{\includegraphics[width=1\textwidth]{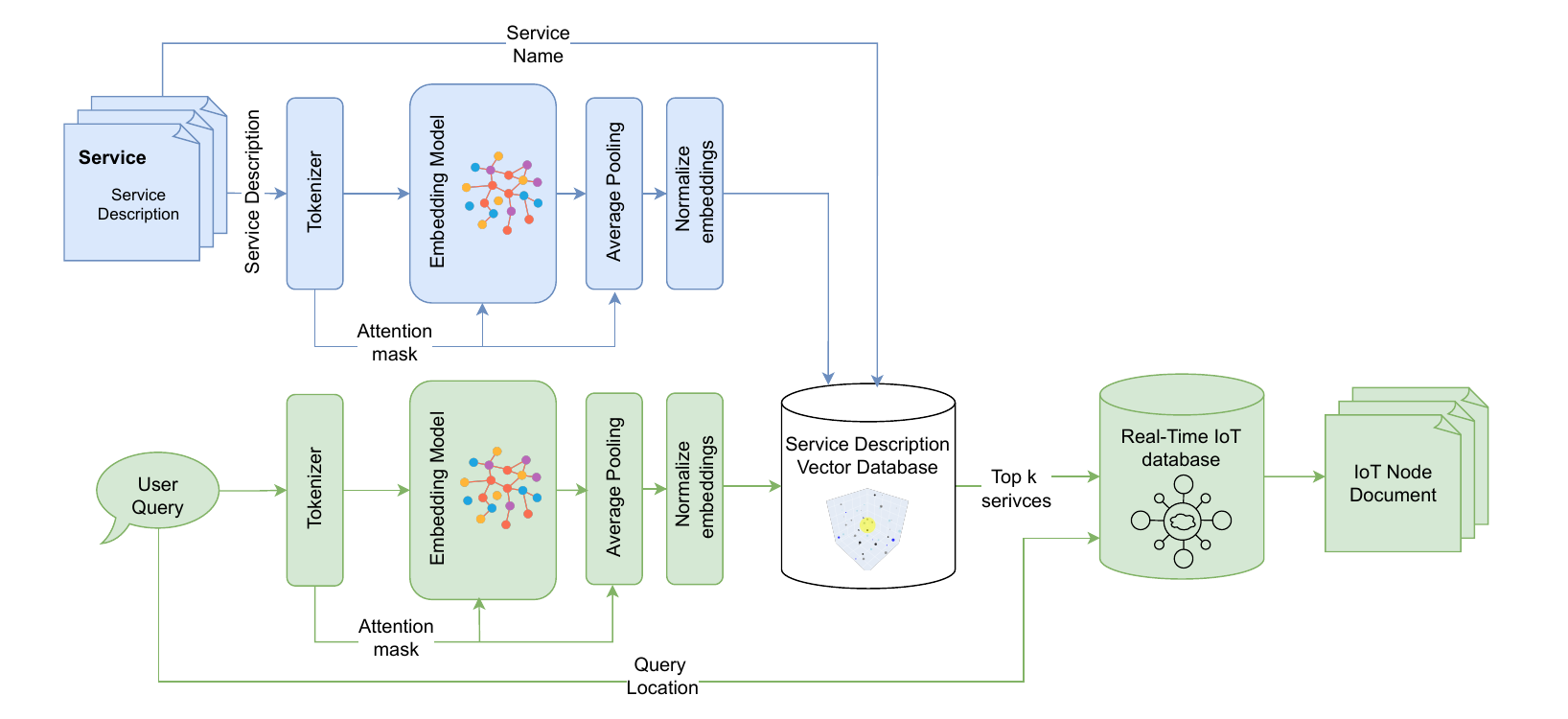}}
\caption{IoT Retrieval-Augmented Generation-Search Engine (IoT-RAG-SE)}
\label{fig:RT-IoT-Engine}
\end{figure*}

\subsection{Generic Agentic RAG (GA-RAG)}
\begin{figure}
\centerline{\includegraphics[width=0.5\textwidth]{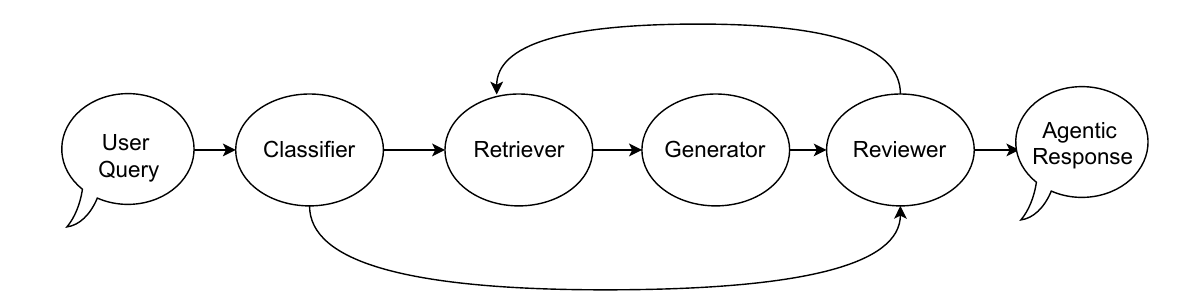}}
\caption{Generic Agentic RAG (GA-RAG) System State Diagram}
\label{fig:state-diagram}
\end{figure}

Fig. \ref{fig:state-diagram} shows the GA-RAG workflow depicted as a state diagram that incorporates four agents: \textit{Classifier}, \textit{Retriever}, \textit{Generator}, and \textit{Reviewer}. The \textit{Classifier} agent determines whether a query is relevant to the GA-RAG services or not. \textit{Retriever} fetches relevant data, mostly containing information needed for response, from the embedded sources (RAG pipeline or other external sources integrated into GA-RAG). \textit{Generator} uses the retrieved context alongside the query and generates the GA-RAG response. Finally, the \textit{Reviewer} agent assesses the generated response before returning the final response in order to reduce the chances of generating inaccurate responses. If the response needs improvement or is unreasonable, it reformulates the query to retrieve a more satisfactory response or direct \textit{Retriever} to use another data source.

\begin{figure}
\begin{frame}
\centering
\animategraphics[autoplay,loop,width=0.5\textwidth]{10}{figures/agentic_system/Image-}{00}{10}
\end{frame}
\caption{IoT Agentic Search Engine (IoT-ASE)}
\label{fig:Agentic-IoT}
\end{figure}
\subsection{Implementation Details}
We used the LangGraph \cite{langchain_langgraph} to build the real-time IoT Agentic Search Engine (IoT-ASE) depicted in Fig. \ref{fig:Agentic-IoT}, based on the Generic Agentic RAG (GA-RAG) workflow shown in Fig. \ref{fig:state-diagram}. We implemented the primary four agents as follows:

\subsubsection{Classifier}  We deploy a virtual assistant agent to classify user queries based on the semantic meaning of the context of the conversation and trigger the adequate \textit{Retriever} node dedicated to the query type. Additionally, if a query is within the LLM capabilities, the virtual assistant agent, as an LLM agent, generates a response and passes it to the reviewer agent. Furthermore, the virtual assistant has access to the region entity explained in the data model, Fig. \ref{fig:data-model}, to know the coverage zone of \textit{IoT-RAG-SE} and use this knowledge in the classification process. Thus, it can decide to route the query to \textit{IoT-RAG-SE} or \textit{GoogleMaps} based on whether the service query is within the SensorsConnect coverage zone or not.
\subsubsection{Retriever} in the \textit{Retriever} node encompasses three retriever sub-nodes:
\begin{enumerate}[label=\roman*]
    \item \textit{IoT-RAG-SE} combines the architecture depicted in Fig. \ref{fig:RT-IoT-Engine}. It has entry to the Service and Node entities described in the data model, Fig. \ref{fig:data-model}  to generate a service descriptions vector database and returns the nearest k node documents, matching user intent, as a context for the Generator node. Additionally, \textit{IoT-RAG-SE} calls OpenRouteService API \cite{openrouteservice} to get travel time and distance matrices between the user location and the places' locations included in the retrieved documents.
    \item \textit{Google Maps} subnode integrates services provided by Google APIs, such as Text Search API. IoT-ASE uses \textit{Google Maps} when a query requests a service unavailable in \textit{IoT-RAG-SE} or a service in an uncovered spatial region. This node returns place documents as a context containing the required details to answer the user query.
    \item The \textit{Scraper} subnode is integrated to handle queries irrelevant to IoT services that LLM can't handle, such as current events and news. It provides documents that may contain the required information based on scraping web pages. \textit{Scraper} uses Tavily \cite{tavily} API to perform searing and scraping process. 
\end{enumerate}
\subsubsection{Generator}
 \textit{Generator} articulates the IoT-ASE response based on the user query and the retrieved context. \textit{Generator} activates particular prompt for each retriever ( i.e. \textit{IoT-RAG-SE}, \textit{Google Maps}, or \textit{Scraper}) to improve the quality of the generative response
\subsubsection{Reviewer}
\textit{Reviewer} checks the user query and the generative response and ensures the response contains the information the user seeks. Additionally, the responses generated by the context of \textit{IoT-RAG-SE} rely on the quality of service descriptions, and inaccurate service descriptions lead to incorrect retrieved contexts. Hence, \textit{Reviewer} can flag the errors propagated from misrepresentation of service descriptions. Also, users may ask for a more precise feature in a service, Such as a query to find an uncrowded Italian restaurant with a discount on family orders and excellent reviews in real-time, so in that case, \textit{Reviewer} ensures the generative reposed addresses these preferences (Italian restaurant, uncrowded, and discount on family orders). For the \textit{Scraper}, \textit{Reviewer} figures out if the answer makes sense only from the meaning. For instance, the generative response should contain a person's name if a user asks a question to know about someone's identity. Thus, {Reviewer} evaluates the semantic meaning of the response and can reroute the user query to another \textit{Retriever} if needed, as shown in the flow graph in Fig. \ref{fig:Agentic-IoT}. Although \textit{Reviewer} has limited capabilities to fully check the correctness of the response, the current architecture tackled the LLM hallucinations problem while developing the prototype version of IoT-ASE. Later on, reviewer abilities can be improved by adding a fact-checking component.

\section{Case Scenario analysis and evaluation}
\label{sec:Use Case}
The introduced \textit{IoT-ASE} can offer services for two entities: (1) the end-users accessing the framework through the user interface layer to boost real-time decision-making, and (2) the IoT systems integrated into the framework to facilitate collaboration between them. In this section, we elaborate on the end-user scenario using a dataset representing real-time IoT sensing data. 

Google has recently launched Gemini \cite{team2023gemini}, a virtual assistant based on LLM models. The Gemini family is a multimodal model capable of understanding texts, images, videos, and audio. Though Gemini \cite{team2023gemini} outperforms other multimodel LLMs and given the simple nature of the IoT data, unlike Human-generated content,  treating it can require a less capable LLM agent and still achieve remarkable performance. Moreover, scraping human-generated content requires computing resources and complex approaches to find the relative content that may contain contexts answering user queries. Using well-structured IoT web content (IoT data hosted in SensorsConnect) would be a more efficient and accurate data source (lightweight contexts), embedding required information sufficient to satisfy user queries. 

In this context, we see the potential of using the introduced \textit{IoT-ASE} in a virtual assistant application that can provide substantial value for the users. Google Maps suggests places based on the user location and ranks the recommended results in ascending order by distance/travel time required to reach there. Although Google Maps estimates occupancy factors of the places, Google Maps has not yet offered a recommendation preference based on service time plus travel (overall time to get served), which in some cases could be efficient in terms of taking times, as we proved before \cite{SensorsConnect}. Furthermore, many preferences can be added to refine the accuracy and relevance of the recommendations. For instance, a user wants to find a park where barbeque is allowed and has an unbooked soccer field. Or, a local restaurant owner preparing to buy the monthly list of stock needs recommendations for wholesale traders with the lowest prices to minimize expenses. This query requires not only keyword matching but also extracting the user's preferences from the user prompt, the personalized component, or the query context. Also, It needs the capability of retrieving IoT data that implies these options so that recommendations based on sophisticated details and criteria can be applicable using the introduced \textit{IoT-ASE}. 

\subsection{Datasets}

\begin{figure}
\centerline{\includegraphics[width=0.5\textwidth]{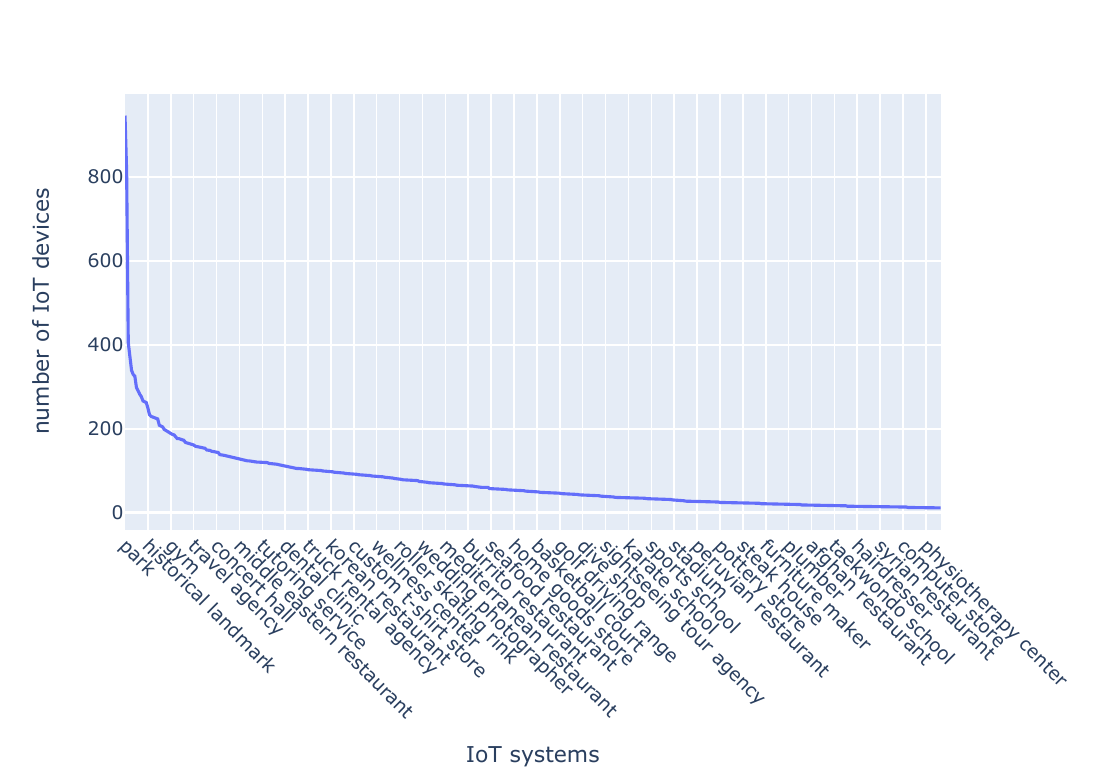}}
\caption{The distribution of IoT devices across a sample of services.}
\label{dataset}
\end{figure}

Google Maps \cite{google_places_api} contains data for more than 200 million places that can mimic IoT systems. We chose the Toronto region to represent the city where we hypothetically deployed IoT-ASE. We built a scraping tool crawling over Google Maps to collect data from the services around the Toronto region that have an online presence on Google Maps. This data is used to simulate IoT devices streaming real-time data. The dataset contains 500 services, which are considered the IoT systems integrated into SensorsConnect. The number of collected instances (IoT devices) included in the evaluation process is 37033. Each service (IoT system) has multiple IoT devices. For example, in the study, parks belong to the park service, and each park represents an instance of an IoT device collecting real-time data. The line graph in Fig. \ref{dataset} depicts the distribution of IoT devices across a sample of services.

The data cleaning process for the collected data includes multiple steps to guarantee accuracy and usability. After collecting raw data, we inspected for any missing values, duplications, or incorrect formats. For some places(instances), the curves of the occupancy factors, which indicate the estimated crowdedness factor of places over opening hours, are missing. Our findings show the places with no occupancy curves are usually uncrowded; hence, their occupancy factors are null.  Also, rates are missing in some instances; given that, most likely, places within the same service type have similar rates, the average rating per service type is the better estimation for the missing rates. 

\subsection{Loading VectorDB and Real-Time IoT Database}
We used the GPT-4 \cite{achiam2023gpt} model to generate service descriptions \footnote{Service descriptions: \url{https://github.com/SensorsConnect/IoT-Agentic-Search-Engine/blob/main/src/vector_db/assets/Services_description_V2.txt}} for the 500 services considered in the case scenario. These descriptions have gone through the embedding process, and then they, alongside their names, are stored and indexed in the service-description vector database.  

Additionally, we stored the dataset containing the scraped services (places) in a MongoDB database in 500 collections, named the same as the names defined in the vector database, so that \textit{IoT-RAG-SE} can route queries based on the requested service and the query location using 2D or Geographical database indexing and return the closest places match the user intent. For example, in the first query listed in Table \ref{table1}, when the \textit{IoT-RAG-SE} agent receives the user query, \textit{"I want to take my dog...."} After returning the top k services \textit{([’dog park’, ’dog walker’, ’dog trainer’])} matching the user query,  the engine uses the returned names of services to locate the collections matching service type. Then, using the service type as a collection name and leveraging Geographical indexing, it can return the nearest places JSON documents (real-time IoT data/context), as shown in the JSON list depicted in Fig. \ref{json:dog_parks}.

\begin{figure}
    \centering
    \begin{lstlisting}[language=json]
[
  {
    "Service Type": "Dog park",
    "Service Name": "Dog Park at Winston Churchill Park",
    "Service Address": "301 St Clair Ave W, Toronto, ON, Canada",
    "Rate": 4.0,
    "Occupancy Factor": 0.50,
    "Travel Time": "7 min"
  },
  {
    "Service Type": "Dog park",
    "Service Name": "Ramsden Dog Park",
    "Service Address": "Toronto, ON, Canada",
    "Rate": 4.8,
    "Occupancy Factor": 0.4,
    "Travel Time": "9 min"
  },
  {
    "Service Type": "Dog park",
    "Service Name": "Cedarvale Park Dogs Off-Leash Area",
    "Service Address": "443 Arlington Ave, York, ON M6C 3A2, Canada",
    "Rate": 4.6,
    "Occupancy Factor": 0.7,
    "Travel Time": "10 min"
  }
]
    \end{lstlisting}
    \caption{Example JSON Data for Dog Parks}
    \label{json:dog_parks}
\end{figure}




\subsection{Evaluation}

\begin{table*}
\begin{tabular}{|l|l|l|l|}
\hline
\textbf{\#} &
  \textbf{intent} &
  \textbf{Query} &
  \textbf{Top k services} \\ \hline
1 &
  Dog Park &
  \begin{tabular}[c]{@{}l@{}}I want to take my dog for walking and playing catch the ball,\\  so I can unleash it\end{tabular} &
  \begin{tabular}[c]{@{}l@{}}{[}'dog park', \\ 'dog walker',\\  'dog trainer'{]}\end{tabular} \\ \hline
2 &
  Shawarma Restaurant &
  \begin{tabular}[c]{@{}l@{}}I'm missing my home country, mmmm! I'm hungry. \\ Oh, I want to eat Shawarma. \\ Can you suggest any nearby place that serves Shawarma?\end{tabular} &
  \begin{tabular}[c]{@{}l@{}}{[}'shawarma restaurant', \\ 'middle eastern restaurant', \\ 'syrian restaurant'{]}\end{tabular} \\ \hline
3 &
  Moving and Storage Service &
  \begin{tabular}[c]{@{}l@{}}I'm planning to move to a new place. \\ Do you know any moving agency close to me?\end{tabular} &
  \begin{tabular}[c]{@{}l@{}}{[}'moving and storage service', \\ 'car rental agency', \\ 'travel agency'{]}\end{tabular} \\ \hline
4 &
  Gym &
  \begin{tabular}[c]{@{}l@{}}I have moved here recently, \\ and I'm looking for a gym with a good reputation.\end{tabular} &
  \begin{tabular}[c]{@{}l@{}}{[}'gym', 'fitness center', \\ 'rock climbing gym'{]}\end{tabular} \\ \hline
5 &
  Car Rental Agency&
  \begin{tabular}[c]{@{}l@{}}I'm travelling tomorrow, and I want to rent a car. \\ Do you know any car rental close to me?\end{tabular} &
  \begin{tabular}[c]{@{}l@{}}{[}'car rental agency', \\ 'vehicle rental agency',\\  'truck rental agency'{]}\end{tabular} \\ \hline
6 &
  Sports School &
  \begin{tabular}[c]{@{}l@{}}My son loves hockey sport,\\  and I want him to start with professional practice playing it. \\ Do you know any hockey school with a good reputation?\end{tabular} &
  \begin{tabular}[c]{@{}l@{}}{[}'sports school', \\ 'hockey club', \\ 'ice skating rink'{]}\end{tabular} \\ \hline
7 &
  Zoo &
  My son loves animals, so I'd like to take him to a zoo &
 \begin{tabular}[c]{@{}l@{}}{[}''zoo', 'wildlife park',\\ 'animal park''{]}\end{tabular} \\ \hline
8 &
  Chinese Restaurant &
  \begin{tabular}[c]{@{}l@{}}I have a conference at Toronto University next week, \\ and I want to have dinner in a Chinese restaurant during my stay there. \\ Can you suggest one with a good reputation?\end{tabular} &
  \begin{tabular}[c]{@{}l@{}}{[}'chinese restaurant', \\ 'canadian restaurant', \\ 'chicken wings restaurant'{]}\end{tabular} \\ \hline
9 &
  Tire Shop &
  \begin{tabular}[c]{@{}l@{}}Winter is coming, and \\ I need to install my winter tires. \\ I'm looking for a place that offers discounts on this service.\end{tabular} &
 \begin{tabular}[c]{@{}l@{}}{[}'auto parts store',\\ 'car detailing service',\\ 'tire shop'{]} \end{tabular} \\ \hline
10 &
  Gift Shop &
  \begin{tabular}[c]{@{}l@{}}My daughter's birthday is next week. \\ Can you suggest a store \\ where I can have a variety of options for her gift?\end{tabular} &
  \begin{tabular}[c]{@{}l@{}}{[}'gift shop', \\ 'toy store', \\ 'souvenir store'{]}\end{tabular} \\ \hline
11 &
  Cocktail Bar &
  \begin{tabular}[c]{@{}l@{}}
It's too hot, and I'm so thirsty.\\ I really want to be hydrated with fresh juice.\\ Do you have any suggestions? \end{tabular}&
  \begin{tabular}[c]{@{}l@{}}{[}['cocktail bar'\\'brunch restaurant'\\, 'bar', ]{]}\end{tabular} \\ \hline
12 &
  Museum &
  \begin{tabular}[c]{@{}l@{}}I'm interested in learning more about the local history. \\ Which museum do you recommend visiting?\end{tabular} &
  \begin{tabular}[c]{@{}l@{}}{[}'memorial park', \\ 'tourist attraction',\\  'museum'{]}\end{tabular} \\ \hline
13 &
  Hair Salon &
  \begin{tabular}[c]{@{}l@{}}I am planning to change my hairstyle\\  and want to visit a top-notch hair salon. \\ Can you suggest a hair salon with a good reputation?\end{tabular} &
  \begin{tabular}[c]{@{}l@{}}{[}'hairdresser', \\ 'hair salon', \\ 'barber shop'{]}\end{tabular} \\ \hline
14 &
  Yoga Center &
  \begin{tabular}[c]{@{}l@{}}I'm stressed these days.\\ Someone told me before that Yoga could help me relieve my stress.\\ Do you have any recommendations for a Yoga center? \end{tabular} &
  \begin{tabular}[c]{@{}l@{}}{[}'yoga center', \\ 'yoga instructor',\\ 'yoga studio'{]}\end{tabular} \\ \hline
15 &
  Furniture Store &
  \begin{tabular}[c]{@{}l@{}}We are redecorating our home and\\  need to find a reliable furniture store with quality products. \\ Can you recommend a furniture store with a good reputation?\end{tabular} &
  \begin{tabular}[c]{@{}l@{}}{[}'antique furniture store', \\ 'furniture accessories',\\  'home goods store'{]}\end{tabular} \\ \hline
16 &
  Dance School &
  \begin{tabular}[c]{@{}l@{}}My daughter loves dancing, and we are looking for a dance school\\  where she can enhance her skills. \\ Can you suggest a dance school with a good reputation?\end{tabular} &
  \begin{tabular}[c]{@{}l@{}}{[}'dance school',\\  'dance company',\\  'music school'{]}\end{tabular} \\ \hline
17 &
  Martial Arts School &
  \begin{tabular}[c]{@{}l@{}}My son is very interested in learning self-defense, \\ and we are looking for a reputable martial arts school.\\  Do you know any martial arts school with a good reputation?\end{tabular} &
  \begin{tabular}[c]{@{}l@{}}{[}'martial arts school', \\ 'karate school', \\ 'taekwondo school'{]}\end{tabular} \\ \hline
18 &
  Medical Spa &
  \begin{tabular}[c]{@{}l@{}}I am planning to treat myself to some relaxation and care,\\  and I am looking for a medical spa with high standards. \\ Do you know any medical spa with a good reputation?\end{tabular} &
  \begin{tabular}[c]{@{}l@{}}{[}'medical spa',\\ 'massage spa',\\  'massage therapist'{]}\end{tabular} \\ \hline
19 &
  Bakery &
  \begin{tabular}[c]{@{}l@{}}My daughter's birthday is coming up, and she loves unique pastries. \\ I'm looking for a bakery that can create a custom cake\\ that's both delicious and visually stunning.\end{tabular} &
  \begin{tabular}[c]{@{}l@{}}{[}'bakery', \\ "children's party service",\\  'donut shop'{]}\end{tabular} \\ \hline
20 &
  Indian Restaurant &
  \begin{tabular}[c]{@{}l@{}}My family and I will visit the local area for a cultural festival. \\ We'd love to experience authentic Indian cuisine while we're there. \\ Could you recommend one nearby with a good reputation?\end{tabular} &
  \begin{tabular}[c]{@{}l@{}}{[}'indian restaurant',\\ 'modern indian restaurant', \\ 'middle eastern restaurant'{]}\end{tabular} \\ \hline
21 &
  Dentist &
  \begin{tabular}[c]{@{}l@{}}My daughter recently had a toothache, and we're looking \\ for a reliable dentist who is experienced with kids ages\\ Could you recommend a good dentist nearby?\end{tabular} &
  \begin{tabular}[c]{@{}l@{}}{[}'dentist', \\ 'dental clinic', \\ 'cosmetic dentist'{]}\end{tabular} \\ \hline
22 &
  Coffee Shop &
  \begin{tabular}[c]{@{}l@{}}I'm planning a casual meeting with a colleague next Tuesday morning. \\ We're looking for a quiet place to discuss some business ideas over coffee. \\ Can you suggest a coffee shop that's known for its serene environment?\end{tabular} &
  \begin{tabular}[c]{@{}l@{}}{[}'coffee shop',\\ 'brunch restaurant', \\ 'lounge'{]}\end{tabular} \\ \hline
23 &
  Optometrist &
  \begin{tabular}[c]{@{}l@{}}My wife has been complaining about her vision while driving at night. \\ We think it might be time for her to see an optometrist. \\ Can you suggest a well-respected optometrist near us?\end{tabular} &
  \begin{tabular}[c]{@{}l@{}}{[}'eye care center', \\ 'optometrist',\\  'optician'{]}\end{tabular} \\ \hline
24 &
  Massage Therapist &
  \begin{tabular}[c]{@{}l@{}}I've been dealing with back pain due to long hours at my desk job. \\ I heard that a good massage can help alleviate some of the pain. \\ Do you know of a massage therapist nearby with excellent reviews?\end{tabular} &
  \begin{tabular}[c]{@{}l@{}}{[}'massage therapist',\\  'massage spa', \\ 'bank'{]}\end{tabular} \\ \hline
25 &
  Golf Club &
  \begin{tabular}[c]{@{}l@{}}It's been a while since I didn't enjoy playing golf. Do you know\\ any nearby golf clubs with affordable membership subscription.\end{tabular} &
  \begin{tabular}[c]{@{}l@{}}{[}'golf club',\\ 'golf shop ', 'golf course'{]}\end{tabular} \\ \hline
\end{tabular}
\vspace{3pt}
\caption{Evaluating RAG stack}
\label{table1}
\end{table*}

We examined the ability of IoT-ASE to 1) comprehend complex queries that imply user preferences embedded in contexts, 2) retrieve documents containing real-time IoT data required to determine the optimum option matches user intent, and 3) generate like-human responses using these retrieved documents as a context. The evaluation involves a list of 25 queries that request services that have service descriptions stored in the vector DB and instances of real-time IoT data stored in the real-time IoT database. Also, we set the user location in the Toronto region unless the user explicitly requests service in a specific city.

\subsubsection{Sematic-Search Evaluation}\hfill\\ 
\textit{IoT-RAG-ASE} shown in Fig. \ref{fig:RT-IoT-Engine} is crucial in the IoT-ASE workflow because it is deployed as a subnode agent to conduct the query understanding task and facilitate real-time context/data in the  IoT-ASE workflow. We conducted separate experiments \cite{Hossam2024} to evaluate the retrieval process of real-time IoT data for several data management systems. Our findings demonstrate that the document data model outperforms other data models in managing real-time IoT data in terms of latency in creating, Reading, Updating, and Deleting (CURD) operations and flexibility in handling heterogeneous data. 
Thus, Highlighting evaluating the query understanding ability, Table \ref{table1} lists the 25 queries to assess the ability of \textit{IoT-RAG-SE} to infer the intended service a user asks via complex queries. The queries were chosen to convey meaning in relatively long contexts containing a subtitle meaning and user preferences to evaluate the performance of the semantic search process discussed in Section \ref{sec:Architecture}. In this experiment, k, the number of top services matching a user query is three, listed in the Top k services column in Table \ref{table1}. For the 25 queries, the engine infers the correct intent in the list of top 3 matched results, with 92 percent of the retrieved results matching the intent in the first result; in queries 9 and 10, Tire Shop and Museum, the engine hits the target intent in the third-ranked result.

Matching the first result is required to minimize the retrieved context by getting the real-time IoT data/documents from the dedicated collection of this service. Nonetheless, to avoid retrieving context from the unintended service, the search engine can retrieve a list of documents from each result, but this approach is inefficient in terms of the cost of token numbers invoking the LLM agent. This evaluation primarily depends on the quality of the service descriptions stored in the vector database. Thus, finding a method for improvement and assessing the services' descriptions can enhance the likelihood of the first matching result.

\subsubsection{Comparison between IoT-ASE and Gemini responses}\hfill\\
 We shed light on three query samples, highlighting contrasts between Gemini and IoT-ASE responses. The prototype \footnote{IoT-ASE: \url{https://github.com/SensorsConnect/IoT-Agentic-Search-Engine}} used to conduct the experiment deploys the llama3-8b-8192 \cite{dubey2024llama} model via Groq \cite{groq2024} API. 

Providing 20 locations, as in Google Maps, sometimes delays making the optimal decision. Similarly, Gemini generates replies that include many preferences, mostly retrieved from Google Maps, leveraging an integrated tool that uses Google Maps API. Seeking to boost decision-making in real-time, we refine the IoT-ASE prompts to generate concise responses, precisely taking into account user preferences when weighing possible options. 

\begin{figure*}
\centerline{\includegraphics[width=0.9\textwidth]{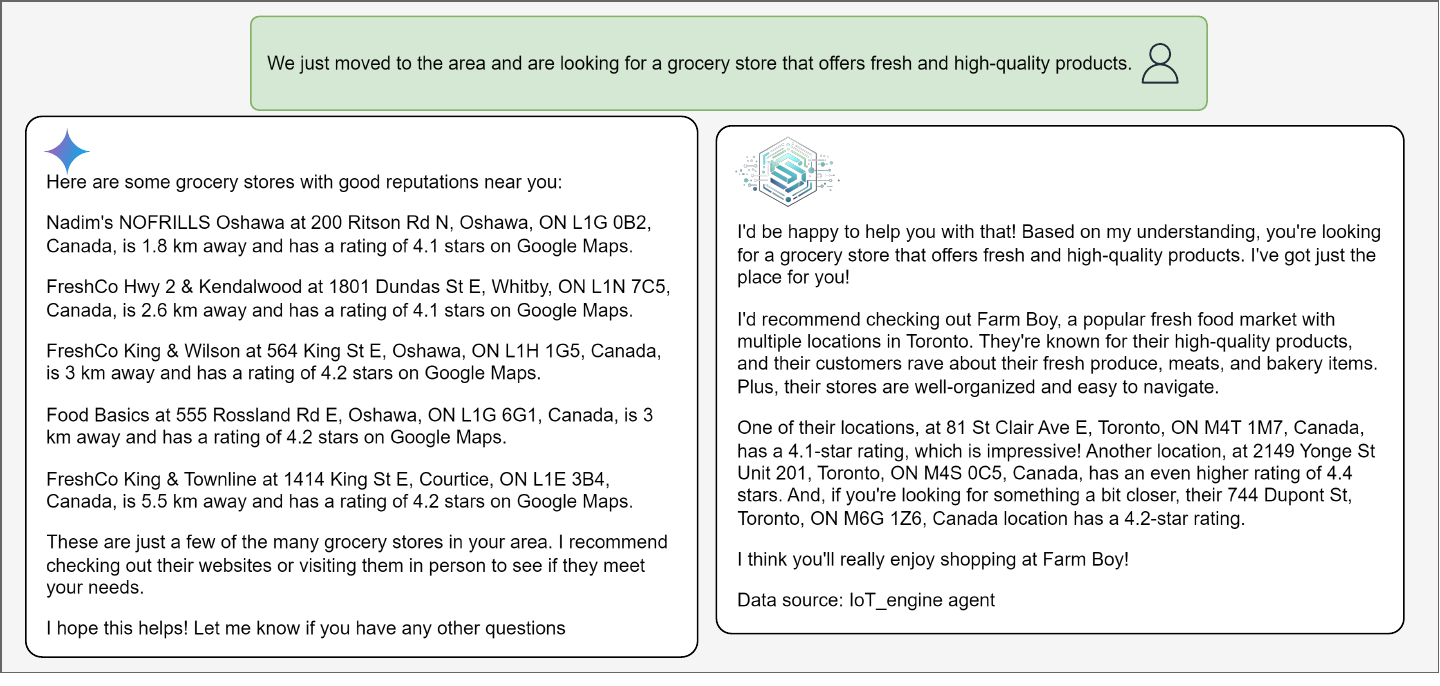}}
\vspace{-14pt}
\caption{Comparison between IoT-ASE and Gemini responses to the Grocery store with a good reputation query, embedding the user location in the context.}
\label{grocery}
\end{figure*}

\begin{figure*}
\centerline{\includegraphics[width=0.9\textwidth]{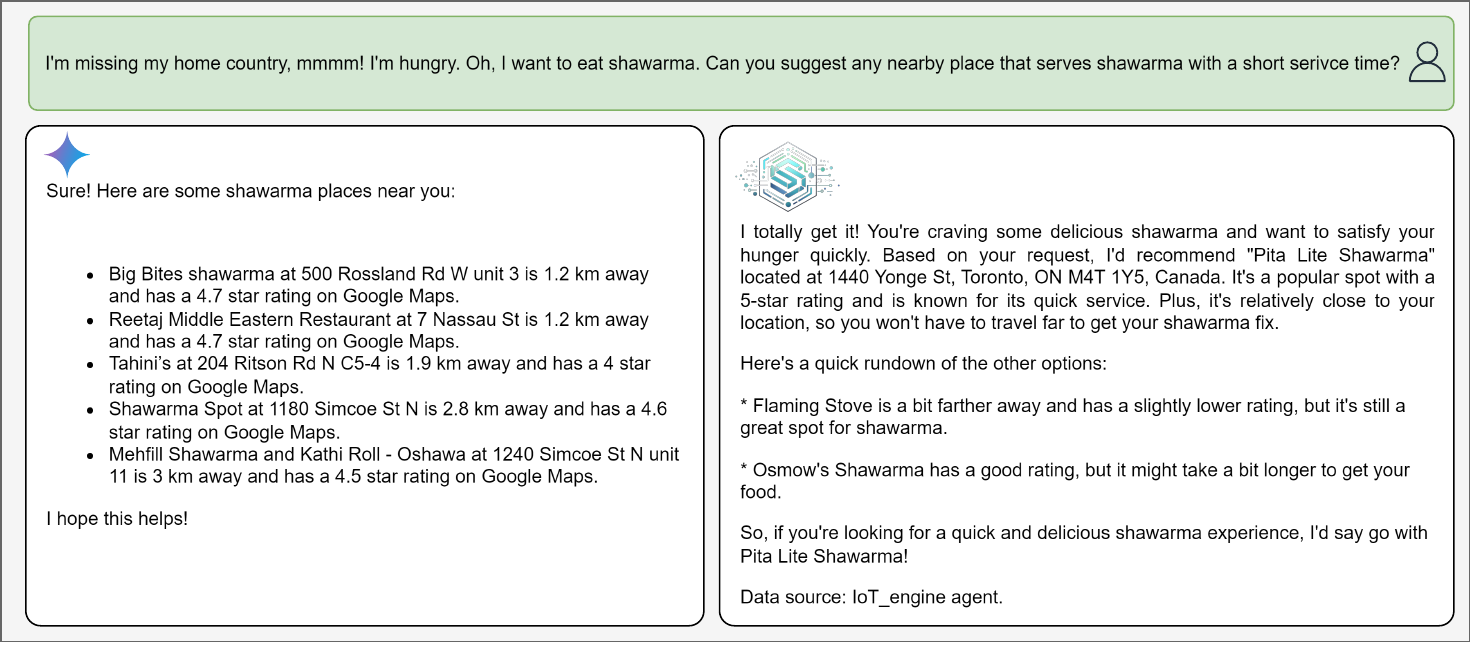}}
\vspace{-14pt}
\caption{Comparison between IoT-ASE and Gemini responses to the Shawarma restaurant with a short service time supported only by IoT-ASE.}
\label{shawrma}
\end{figure*}

\begin{figure*}
\centerline{\includegraphics[width=0.9\textwidth]{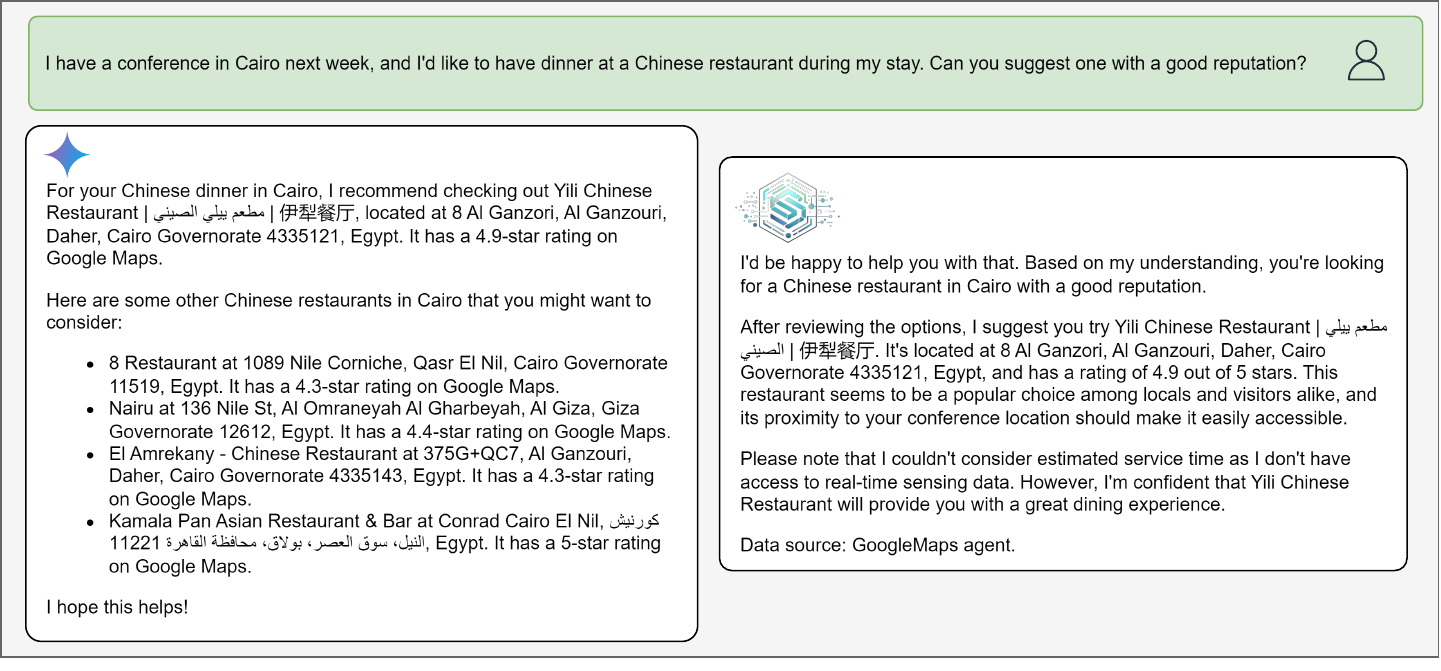}}
\vspace{-14pt}
\caption{Comparison between IoT-ASE and Gemini responses to the Chinese restaurant in Cairo, highlighting recognizing the uncovered zone.}
\label{Chinese}
\end{figure*}

For instance,  in Fig. \ref{grocery}, \ref{shawrma}, and \ref{Chinese}, Gemini provides responses containing several choices, listed similarly to Google Maps results but in bullet points and, in contrast, the IoT-ASE produces humanized responses that recommend a single option after comparing three potential options based on user preferences. Furthermore, these three samples reveal that IoT-ASE responses are more relevant and concise, accurately addressing the implied preferences. However, Gemini's responses follow a generalized pattern and do not indicate if these options accurately match the implied preferences in the contexts. Compared to the generalized Gemini response in the grocery query in Fig. \ref{grocery}, a user is looking for a grocery store with a good reputation. IoT-ASE inferences the service the user is looking for, embedded preferences like a good reputation, and the service location, and uses this knowledge to search for services close to the required location and determine the closest k services. After retrieving k documents of real-time IoT data, IoT-ASE weighed the options based on preferences embedded in the query context. Finally, it generated a persuasive reply by embedding phrases like fresh and high-quality products,  a popular fresh market, and their customer rave about their fresh food, in addition to the ability to use the provided distances and get travel times using OpenRouteService API to navigate the user to the closest branch. 

In the Shawarma query depicted in Fig. \ref{shawrma}, regardless of whether it is a generalized or precise response based on user preferences, Gemini cannot access real-time sensing data about service time. Conversely, IoT-ASE leverages retrieving the occupancy factors from the real-time IoT database alongside travel time using the OpenRouteService API. Thus, IoT-ASE can generate a more precise response that addresses user preferences, leveraging real-time IoT data, as shown in this example. 

In the Chinese restaurant example, Fig. \ref{Chinese}, IoT-ASE responds using a context retrieved from the GoogleMaps agent because the study includes places in the Toronto region, so Cairo is uncovered by the \textit{IoT-RAG-SE}. In other words, the \textit{Virtual Assistant} agent in IoT-ASE, Fig. \ref{fig:Agentic-IoT},  recognized Cairo as an uncovered zone because it is not registered in the coverage zone database shown in the diagram, so \textit{Virtual Assistant}  routed the query to the \textit{Google Maps} agent. Even though, IoT-ASE used context-retrieved
from Google Maps, it generated a more engaged response than the one generated by Gemini. There is no doubt Gemini outperforms IoT-ASE in responding to general prompts and the other broad range of tasks. Nonetheless, In this study, we try to show that the introduced architecture is applicable and can provide LLMs with not only a real-time data source, boosting real-time decision-making but also lightweight contexts instead of resorting to scraping human content (the WWW content), which requires more computing resources.

\subsubsection{discussions}
\begin{figure}
\centerline{\includegraphics[width=0.5\textwidth]{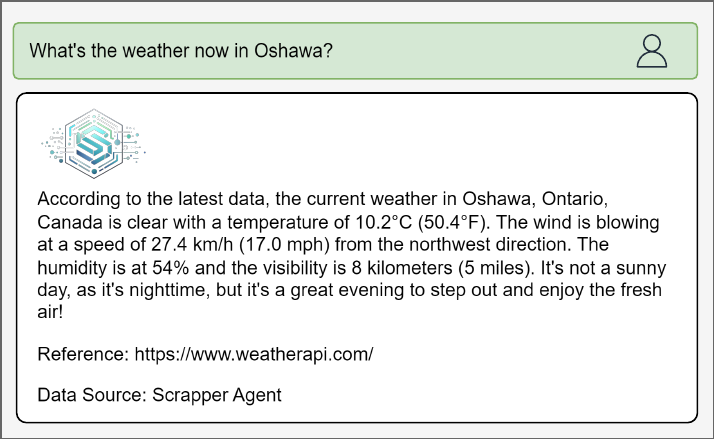}}
\vspace{-10pt}
\caption{The response of IoT-ASE to the weather in Oshawa query by \textit{scraping} agent that uses the weather API. }
\label{weather}
\end{figure}

\hfill\\
LLM agents can rely on a scraping approach to find real-time IoT data that can answer user queries, but this process requires an intensive resource. For instance, in Fig. \ref{weather}, a user queries the weather status in Oshawa. Because IoT-ASE does not support the weather service, IoT-ASE routes the query to the scraping agent, which uses Tavily  API to scrape and provide the relative context containing the weather conditions in Oshawa. Fig. \ref{weather} shows the retrieved Tavily content, which surprisingly is not collected from a scraping process. It uses Weatherapi \cite{weatherapi2024}, which provides APIs for real-time weather conditions. In other words, Tavily started implementing particular tools from some queries, as this example definitely shows the integration of the Weather condition tool so that weather queries can be efficiently handled instead of carrying out a scraping process. However, building or integrating endless APIs to leverage real-time IoT data for myriad IoT systems is an infeasible solution. Therefore, IoT-ASE leverages a unified data model, Fig. \ref{fig:data-model}, to circumvent data scraping or integrating endless APIs.

Furthermore, We collected brief statistics via LangSmith \cite{langsmith2024} tracing tools. For 100 queries, IoT-ASE consumed 126,769 tokens with a median of 1330. Among these queries, there was a 1\% error rate that IoT-ASE could not address. In addition, 50 percent of the responses are processed in less than 2.10 seconds, and 99\% are generated within 4.12 seconds. Hence, considering the nature of the use-case scenario, these statistics indicate that IoT-ASE can maintain an acceptable range of unhandled queries resulting from internal errors occurring within IoT-ASE and achieve relatively low latency, which is critical in dealing with real-time applications. 

\begin{figure*}
\centerline{\includegraphics[width=1\textwidth]{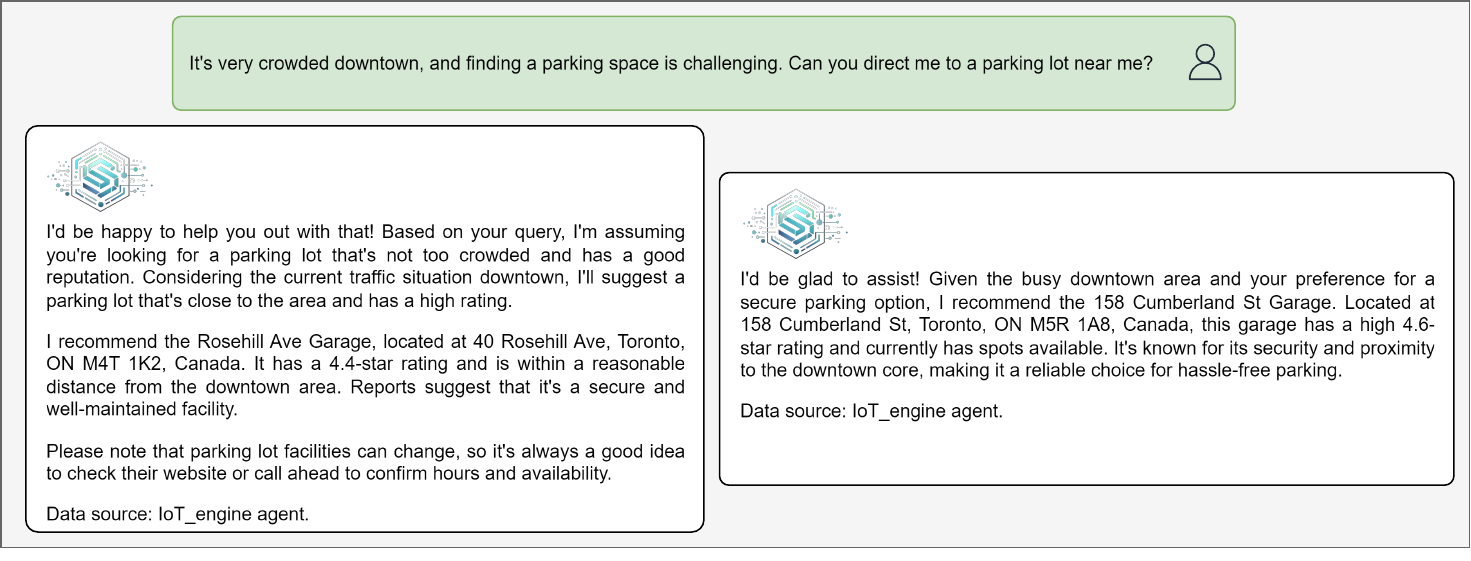}}
\caption{Comparison of IoT-ASE responses with and without the available parking spots field.}
\label{parking-lot}
\end{figure*}

\begin{figure*}
\centerline{\includegraphics[width=1\textwidth]{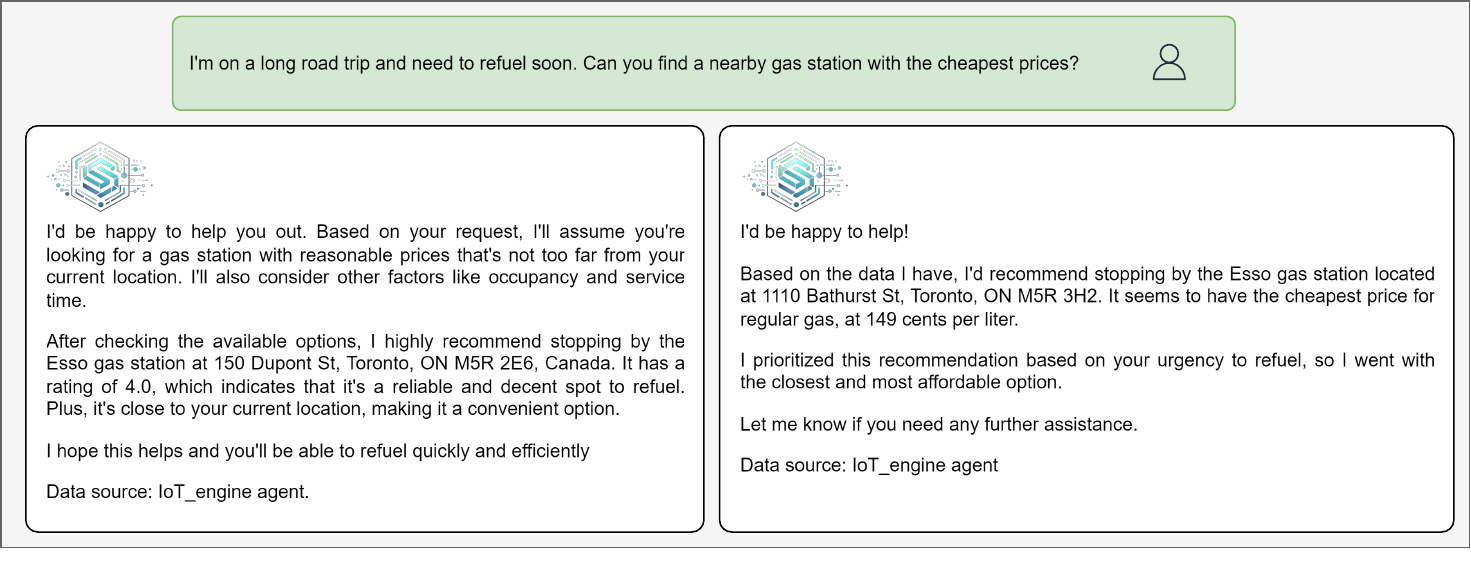}}
\caption{Comparison of IoT-ASE responses with and without the gas price field.}
\label{gas-station}
\end{figure*}

\begin{figure*}
\centerline{\includegraphics[width=1\textwidth]{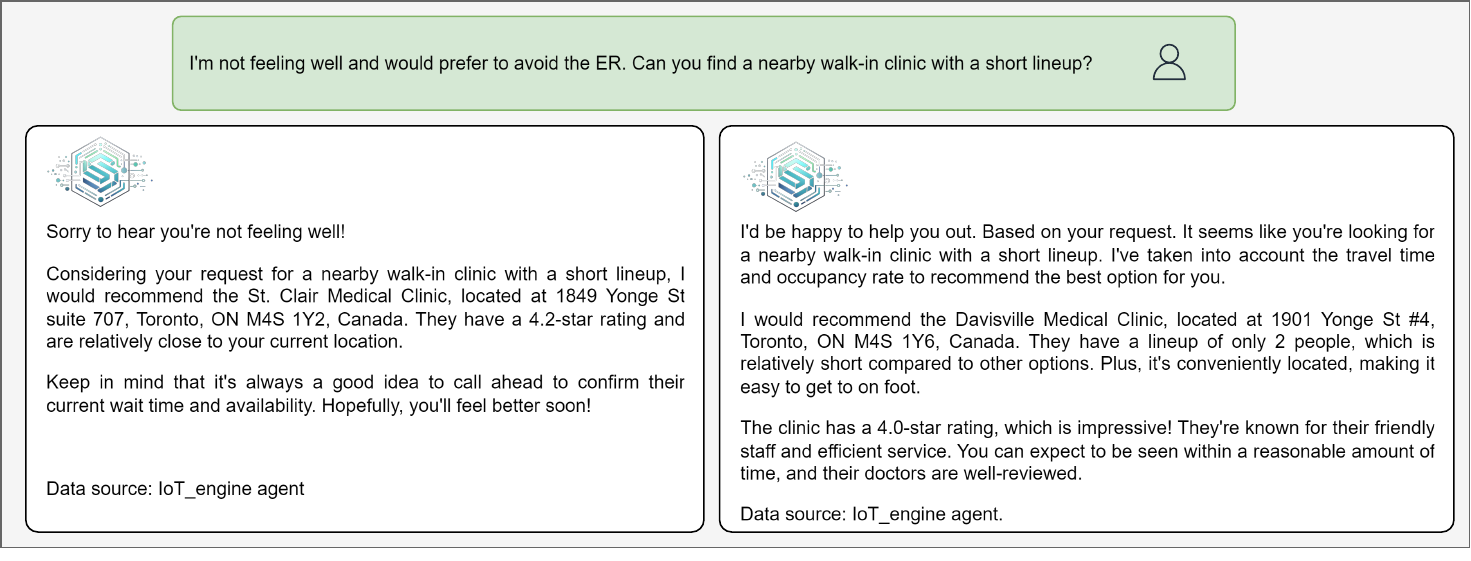}}
\caption{Comparison of IoT-ASE responses with and without the lineup counter field}
\label{walk-in}
\end{figure*}

Lastly,  Figures \ref{parking-lot}, \ref{gas-station}, and \ref{walk-in} present scenarios showing the potential impact of IoT-ASE and illuminating the merits of having access to real-time IoT. SensorsConnect \cite{SensorsConnect} authorizes service providers (places) to update their real-time IoT data instantly. Fig. \ref{parking-lot} depicts a scenario where a user is located downtown during rush hour, driving around to find a parking spot. The response on the left side was generated without including an available parking spot field for every retrieved document, which the generator agent uses as a context to decide the optimum choice. Then, we updated these documents by including a boolean-type field showing the status of parking spot availability. We intentionally presumed the suggested choice in the previous response was fully occupied to see if the generator agent could decide based on the parking spot availability field. After updating the documents, IoT-ASE assessed the given documents of garages and successfully improved the recommendation quality by suggesting the one with the available parking spots.

Additionally, Fig. \ref{gas-station} illustrates another query in which a user was looking for a gas station with the cheapest price. Similarly, We assessed IoT-ASE using the same approach demonstrated in the parking lot scenario. Without having the gas price field, IoT-ASE is recommended based on other factors like the travel distance, rate, and occupancy factor, and given the user needs to refuel his vehicle soon, so IoT-ASE weight the other factors to help the user to decide without hesitating as occurring Gemini or Google Maps. On the other hand, in the response on the right side, IoT-ASE compares three prices listed in the retrieved documents of gas stations near the user location and recommends, based only on the user performance, the cheapest choice among the close gas stations.

Recent statistics \cite{Walk_in2023wait} show that the average wait time of a patient in the Walk-in clinics in Toronto is about 72 minutes. Utilizing real-time data on expected wait times in clinics, IoT-ASE can assist healthcare providers in distributing patients across clinics based on current occupancy levels. Fig. \ref{walk-in} exhibits two IoT-ASE responses for a user query for a walk-in clinic with a short lineup within his current location. IoT-ASE provided the answer on the left side without embedding the lineup field in the clinics' documents in the real-time IoT database. IoT-ASE suggested that the user call the clinic to ensure availability and to know the expected wait time. In contrast, by embedding the current lineup field in clinics' documents, IoT-ASE could consider it to elect the clinic with the shortest lineup. Thus, deploying IoT-ASE can help mitigate the long wait time of patients and evenly coordinate patients between healthcare providers.

Still, there are many scenarios and fields that can leverage updating and accessing real-time IoT data. For instance, for industries in which their products have short expiration dates, e.g. in the food industry, around 12\% \cite{canada_food_waste_2024} of food sales in retail were reported as food waste in Canada during 2019. IoT-ASE can connect customers to stores, applying offers and sales on products that will expire soon to reach the customers they are looking for. Therefore, these scenarios prove that IoT-ASE can elevate decision-making in real-time and, consequently, quality of life.

\section{Conclusion}
\label{sec:Conclusion}
The paper proposes an \textit{IoT-RAG-SE} using the Retrieval-Augmented Generation (RAG) stack that processes two tasks: embedding service descriptions and performing semantic searches between user queries and service descriptions to retrieve real-time IoT data from real-time IoT database without embedding IoT data. Furthermore, the paper defines the Generic Agentic-RAG (GA-RAG) workflow by analyzing existing agentic systems and evaluating the requirements of RAG architectures. It also presents the implementation of IoT Agentic Search Engine (IoT-ASE) workflow based on GA-RAG. 

We demonstrate the usability and performance of the proposed search engine with a use-case scenario where IoT-ASE is used as a virtual assistant that can refine service recommendations, leveraging the availability of real-time IoT data. We hypothetically deploy IoT-ASE in the Toronto region with a dataset that combines 500 services mimicking IoT systems and 37033 JSON documents of places mimicking IoT devices. The evaluation process combines a set of 25 complex queries that embed user preferences in the context to evaluate the performance of \textit{IoT-RAG-SE} to understand the user intent. The statistics show that \textit{IoT-RAG-SE} retrieves the intent service as the top result with an accuracy of 92\%. Also, the evaluation sheds light on a sample of queries that assess and evaluate the performance of IoT-ASE compared to Gemini. Our investigations conclude that IoT-ASE is able to generate more accurate and precise responses, addressing user preferences. In contrast, Gemini tends to generalize the responses without fully addressing preferences embedded in the queries. Even though, theoretically, scraping can be used to find IoT data over WWW or connect endless tools that provide IoT data as context to LLM, relying on SensorsConnect, enabling the use of the unified data model defined in the paper, eases the process of getting lightweight context containing the intended data. Based on our evaluation and discussion, we consider IoT-ASE to be a promising solution that enables the accessibility of real-time IoT data and boosts real-time decision-making.

\ifCLASSOPTIONcaptionsoff
  \newpage
\fi

\clearpage
\vspace{-10pt}
\bibliographystyle{IEEEtran}
\bibliography{bibtex/bib/main}

\begin{thebibliography}{10}
\providecommand{\url}[1]{#1}
\csname url@samestyle\endcsname
\providecommand{\newblock}{\relax}
\providecommand{\bibinfo}[2]{#2}
\providecommand{\BIBentrySTDinterwordspacing}{\spaceskip=0pt\relax}
\providecommand{\BIBentryALTinterwordstretchfactor}{4}
\providecommand{\BIBentryALTinterwordspacing}{\spaceskip=\fontdimen2\font plus
\BIBentryALTinterwordstretchfactor\fontdimen3\font minus \fontdimen4\font\relax}
\providecommand{\BIBforeignlanguage}[2]{{%
\expandafter\ifx\csname l@#1\endcsname\relax
\typeout{** WARNING: IEEEtran.bst: No hyphenation pattern has been}%
\typeout{** loaded for the language `#1'. Using the pattern for}%
\typeout{** the default language instead.}%
\else
\language=\csname l@#1\endcsname
\fi
#2}}
\providecommand{\BIBdecl}{\relax}
\BIBdecl

\bibitem{explodingtopics2024iot}
\BIBentryALTinterwordspacing
E.~Topics, ``Number of iot devices in 2024,'' 2024, accessed: 2024-07-15. [Online]. Available: \url{https://explodingtopics.com/blog/number-of-iot-devices}
\BIBentrySTDinterwordspacing

\bibitem{elgazzar2022revisiting}
K.~Elgazzar, H.~Khalil, T.~Alghamdi, A.~Badr, G.~Abdelkader, A.~Elewah, and R.~Buyya, ``Revisiting the internet of things: New trends, opportunities and grand challenges,'' p. 1073780, 2022.

\bibitem{IoTSearch1}
J.~Han, L.~Qi, and J.~Zhuang, ``Vector sum range decision for verifiable multiuser fuzzy keyword search in cloud-assisted iot,'' \emph{IEEE Internet of Things Journal}, vol.~11, no.~1, pp. 931--943, 2024.

\bibitem{IoTSearch2}
W.~G. Hatcher, C.~Qian, F.~Liang, W.~Liao, E.~P. Blasch, and W.~Yu, ``Secure iot search engine: Survey, challenges issues, case study, and future research direction,'' \emph{IEEE Internet of Things Journal}, vol.~9, no.~18, pp. 16\,807--16\,823, 2022.

\bibitem{shodan}
\BIBentryALTinterwordspacing
{Shodan}, ``{Shodan: Search Engine for the Internet of Everything},'' 2024, [Accessed: 14-November-2024]. [Online]. Available: \url{https://www.shodan.io/}
\BIBentrySTDinterwordspacing

\bibitem{mulero2023detection}
S.~Mulero-Palencia and V.~Monzon~Baeza, ``Detection of vulnerabilities in smart buildings using the shodan tool,'' \emph{Electronics}, vol.~12, no.~23, p. 4815, 2023.

\bibitem{liang2019search}
F.~Liang, C.~Qian, W.~G. Hatcher, and W.~Yu, ``Search engine for the internet of things: Lessons from web search, vision, and opportunities,'' \emph{IEEE Access}, vol.~7, pp. 104\,673--104\,691, 2019.

\bibitem{censys}
\BIBentryALTinterwordspacing
{Censys}, ``{The Censys Platform: The One Place to Understand Everything on the Internet},'' 2024, [Accessed: 14-November-2024]. [Online]. Available: \url{https://censys.com/}
\BIBentrySTDinterwordspacing

\bibitem{crowdstrike2024}
M.~Sentonas, ``Crowdstrike to acquire reposify to reduce risk across the external attack surface and fortify customer security postures,'' \url{https://www.crowdstrike.com/blog/crowdstrike-to-acquire-reposify-to-reduce-risk-across-the-external-attack-surface-and-fortify-customer-security-postures/}, September 2022, [Online; accessed 20-July-2024].

\bibitem{iggena2021iotcrawler}
T.~Iggena, E.~Bin~Ilyas, M.~Fischer, R.~T{\"o}njes, T.~Elsaleh, R.~Rezvani, N.~Pourshahrokhi, S.~Bischof, A.~Fernbach, J.~X. Parreira \emph{et~al.}, ``Iotcrawler: Challenges and solutions for searching the internet of things,'' \emph{Sensors}, vol.~21, no.~5, p. 1559, 2021.

\bibitem{elsaleh2020iot}
T.~Elsaleh, S.~Enshaeifar, R.~Rezvani, S.~T. Acton, V.~Janeiko, and M.~Bermudez-Edo, ``Iot-stream: A lightweight ontology for internet of things data streams and its use with data analytics and event detection services,'' \emph{Sensors}, vol.~20, no.~4, p. 953, 2020.

\bibitem{SensorsConnect}
A.~Elewah and K.~Elgazzar, ``Sensorsconnect framework: World-wide web for internet of things,'' \emph{IEEE Access}, pp. 1--1, 2024.

\bibitem{brin1998anatomy}
S.~Brin and L.~Page, ``The anatomy of a large-scale hypertextual web search engine,'' \emph{Computer networks and ISDN systems}, vol.~30, no. 1-7, pp. 107--117, 1998.

\bibitem{spatharioti2023comparing}
S.~E. Spatharioti, D.~M. Rothschild, D.~G. Goldstein, and J.~M. Hofman, ``Comparing traditional and llm-based search for consumer choice: A randomized experiment,'' \emph{arXiv preprint arXiv:2307.03744}, 2023.

\bibitem{minaee2024large}
S.~Minaee, T.~Mikolov, N.~Nikzad, M.~Chenaghlu, R.~Socher, X.~Amatriain, and J.~Gao, ``Large language models: A survey,'' \emph{arXiv preprint arXiv:2402.06196}, 2024.

\bibitem{lewis2020retrieval}
P.~Lewis, E.~Perez, A.~Piktus, F.~Petroni, V.~Karpukhin, N.~Goyal, H.~K{\"u}ttler, M.~Lewis, W.-t. Yih, T.~Rockt{\"a}schel \emph{et~al.}, ``Retrieval-augmented generation for knowledge-intensive nlp tasks,'' \emph{Advances in Neural Information Processing Systems}, vol.~33, pp. 9459--9474, 2020.

\bibitem{berners1994world}
T.~Berners-Lee, R.~Cailliau, A.~Luotonen, H.~F. Nielsen, and A.~Secret, ``The world-wide web,'' \emph{Communications of the ACM}, vol.~37, no.~8, pp. 76--82, 1994.

\bibitem{elewah2022thingsdriver}
A.~Elewah, W.~M. Ibrahim, A.~Raf{\i}kl, and K.~Elgazzar, ``Thingsdriver: A unified interoperable driver for iot nodes,'' in \emph{2022 International Wireless Communications and Mobile Computing (IWCMC)}.\hskip 1em plus 0.5em minus 0.4em\relax IEEE, 2022, pp. 877--882.

\bibitem{Traffic}
S.~Mostafi, T.~Alghamdi, and K.~Elgazzar, ``Interconnected traffic forecasting using time distributed encoder-decoder multivariate multi-step lstm,'' in \emph{2024 IEEE Intelligent Vehicles Symposium (IV)}, 2024, pp. 2503--2508.

\bibitem{ServicesDiscovery}
K.~Elgazzar, A.~E. Hassan, and P.~Martin, ``Clustering wsdl documents to bootstrap the discovery of web services,'' in \emph{2010 IEEE international conference on web services}.\hskip 1em plus 0.5em minus 0.4em\relax IEEE, 2010, pp. 147--154.

\bibitem{elgazzar2014daas}
K.~Elgazzar, H.~S. Hassanein, and P.~Martin, ``Daas: Cloud-based mobile web service discovery,'' \emph{Pervasive and Mobile Computing}, vol.~13, pp. 67--84, 2014.

\bibitem{vaswani2017attention}
A.~Vaswani, N.~Shazeer, N.~Parmar, J.~Uszkoreit, L.~Jones, A.~N. Gomez, {\L}.~Kaiser, and I.~Polosukhin, ``Attention is all you need,'' \emph{Advances in neural information processing systems}, vol.~30, 2017.

\bibitem{achiam2023gpt}
J.~Achiam, S.~Adler, S.~Agarwal, L.~Ahmad, I.~Akkaya, F.~L. Aleman, D.~Almeida, J.~Altenschmidt, S.~Altman, S.~Anadkat \emph{et~al.}, ``Gpt-4 technical report,'' \emph{arXiv preprint arXiv:2303.08774}, 2023.

\bibitem{li2022survey}
H.~Li, Y.~Su, D.~Cai, Y.~Wang, and L.~Liu, ``A survey on retrieval-augmented text generation,'' \emph{arXiv preprint arXiv:2202.01110}, 2022.

\bibitem{zhao2024retrieval}
P.~Zhao, H.~Zhang, Q.~Yu, Z.~Wang, Y.~Geng, F.~Fu, L.~Yang, W.~Zhang, and B.~Cui, ``Retrieval-augmented generation for ai-generated content: A survey,'' \emph{arXiv preprint arXiv:2402.19473}, 2024.

\bibitem{yao2022react}
S.~Yao, J.~Zhao, D.~Yu, N.~Du, I.~Shafran, K.~Narasimhan, and Y.~Cao, ``React: Synergizing reasoning and acting in language models,'' \emph{arXiv preprint arXiv:2210.03629}, 2022.

\bibitem{venkatraman2024collabstory}
S.~Venkatraman, N.~I. Tripto, and D.~Lee, ``Collabstory: Multi-llm collaborative story generation and authorship analysis,'' \emph{arXiv preprint arXiv:2406.12665}, 2024.

\bibitem{li2023tradinggpt}
Y.~Li, Y.~Yu, H.~Li, Z.~Chen, and K.~Khashanah, ``Tradinggpt: Multi-agent system with layered memory and distinct characters for enhanced financial trading performance,'' \emph{arXiv preprint arXiv:2309.03736}, 2023.

\bibitem{berenguer2024leveraging}
A.~Berenguer, A.~Morej{\'o}n, D.~Tom{\'a}s, and J.-N. Maz{\'o}n, ``Leveraging large language models for sensor data retrieval,'' \emph{Applied Sciences}, vol.~14, no.~6, p. 2506, 2024.

\bibitem{Hossam2024}
H.~Ouda, A.~Elewah, and K.~Elgazzar, ``A comparative analysis of data models for heterogeneous sensor data management,'' in \emph{2024 International Wireless Communications and Mobile Computing (IWCMC)}, 2024, pp. 1826--1833.

\bibitem{reimers2019sentence}
N.~Reimers, ``Sentence-bert: Sentence embeddings using siamese bert-networks,'' \emph{arXiv preprint arXiv:1908.10084}, 2019.

\bibitem{malkov2018efficient}
Y.~A. Malkov and D.~A. Yashunin, ``Efficient and robust approximate nearest neighbor search using hierarchical navigable small world graphs,'' \emph{IEEE transactions on pattern analysis and machine intelligence}, vol.~42, no.~4, pp. 824--836, 2018.

\bibitem{langchain_langgraph}
LangChain, ``Balance agent control with agency,'' \url{https://www.langchain.com/langgraph}, accessed: 2024-9-17.

\bibitem{openrouteservice}
OpenRouteService, ``Openrouteservice api services,'' \url{https://openrouteservice.org/}, accessed: 2024-8-17.

\bibitem{tavily}
Tavily, ``Connect your llm to the web,'' \url{https://tavily.com/}, accessed: 2024-8-17.

\bibitem{team2023gemini}
G.~Team, R.~Anil, S.~Borgeaud, Y.~Wu, J.-B. Alayrac, J.~Yu, R.~Soricut, J.~Schalkwyk, A.~M. Dai, A.~Hauth \emph{et~al.}, ``Gemini: a family of highly capable multimodal models,'' \emph{arXiv preprint arXiv:2312.11805}, 2023.

\bibitem{google_places_api}
\BIBentryALTinterwordspacing
{Google Developers}, ``{Places API Web Service Documentation},'' 2024, [Accessed: 18-May-2024]. [Online]. Available: \url{https://developers.google.com/maps/documentation/places/web-service}
\BIBentrySTDinterwordspacing

\bibitem{dubey2024llama}
A.~Dubey, A.~Jauhri, A.~Pandey, A.~Kadian, A.~Al-Dahle, A.~Letman, A.~Mathur, A.~Schelten, A.~Yang, A.~Fan \emph{et~al.}, ``The llama 3 herd of models,'' \emph{arXiv preprint arXiv:2407.21783}, 2024.

\bibitem{groq2024}
\BIBentryALTinterwordspacing
{Groq, Inc.}, ``{Groq: Fast AI Inference},'' 2024, [Accessed: 20-August-2024]. [Online]. Available: \url{https://groq.com/}
\BIBentrySTDinterwordspacing

\bibitem{weatherapi2024}
\BIBentryALTinterwordspacing
{WeatherAPI.com}, ``{Real Time, Forecasted, Future, Marine and Historical Weather},'' 2024, [Accessed: 20-August-2024]. [Online]. Available: \url{https://www.weatherapi.com/}
\BIBentrySTDinterwordspacing

\bibitem{langsmith2024}
\BIBentryALTinterwordspacing
{LangChain, Inc.}, ``{Get your LLM app from prototype to production},'' 2024, [Accessed: 20-August-2024]. [Online]. Available: \url{https://www.langchain.com/langsmith}
\BIBentrySTDinterwordspacing

\bibitem{Walk_in2023wait}
\BIBentryALTinterwordspacing
medimap Team, ``The walk-in clinic wait time index,'' Medimap, Canada, Tech. Rep., 2024, data collected from walk-in clinics across Canada to analyze average wait times by province. [Online]. Available: \url{https://medimap.ca}
\BIBentrySTDinterwordspacing

\bibitem{canada_food_waste_2024}
Environment and C.~C. Canada, ``Taking stock: Reducing food loss and waste in canada,'' \url{https://www.canada.ca/en/environment-climate-change/services/managing-reducing-waste/food-loss-waste/taking-stock.html}, 2024, accessed: 2024-10-31.

\end{thebibliography}

\begin{IEEEbiography}
[{\includegraphics[width=1in,height=1.25in,clip,keepaspectratio]{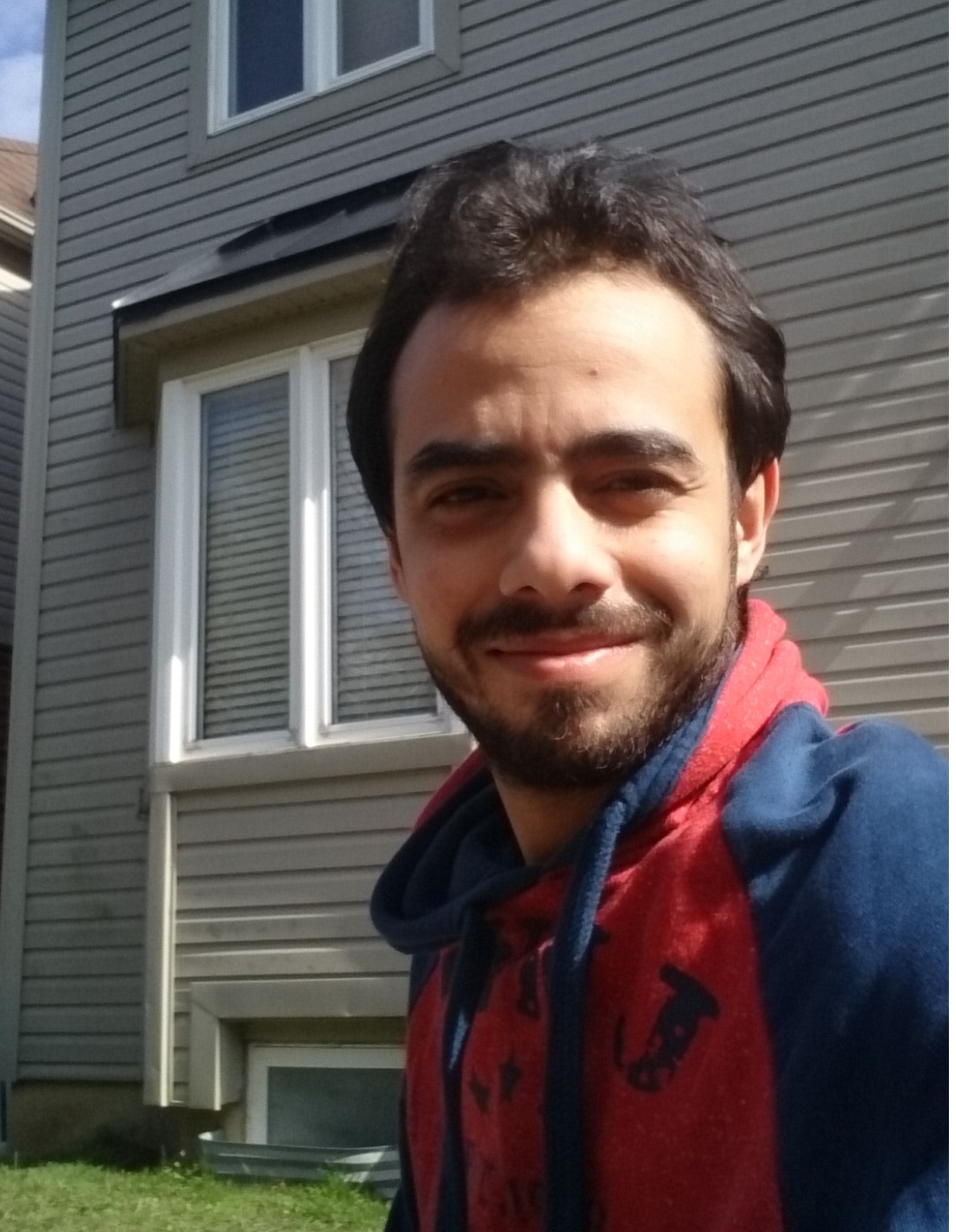}}]{Abdelrahman Elewah} is a Ph.D. candidate at Ontario Tech University and a member of the IoT Research Lab. He earned his master's degree in Electrical and Electronics Engineering and his bachelor's degree in Technology, Electronics, and Communications Engineering from Benha University, Egypt. His research focuses on IoT, ML, and Generative AI landscapes through SensorsConnect, a framework designed to enable IoT systems to interact seamlessly, much like the human experience of the World Wide Web.
\end{IEEEbiography}
\begin{IEEEbiography}[{\includegraphics[width=1in,height=1.25in,clip,keepaspectratio]{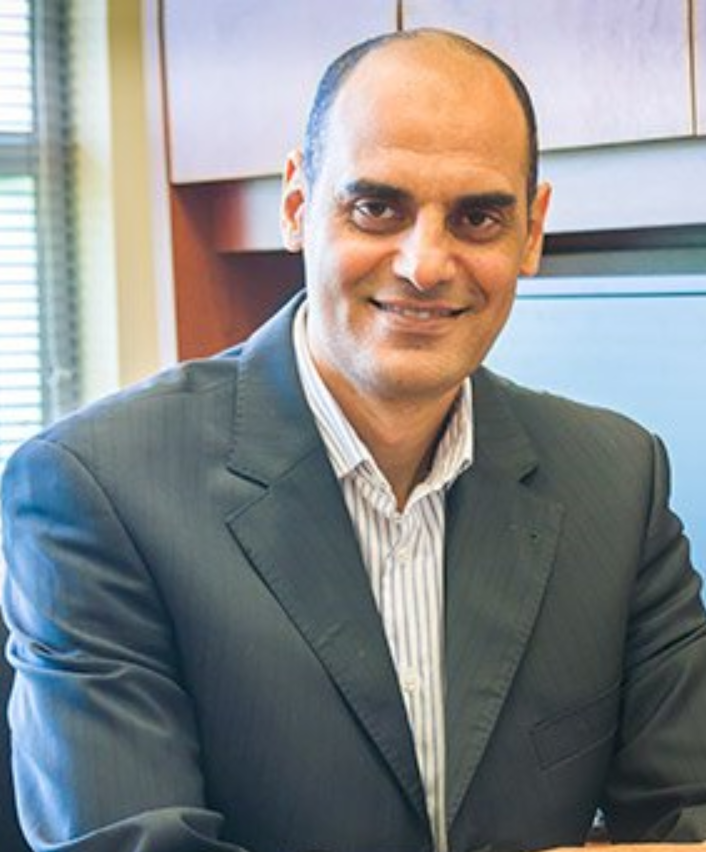}}]{Dr. Khalid Elgazzar} 
is a Canada Research Chair and Associate Professor in the Faculty of Engineering and Applied Science at Ontario Tech University, Canada, and he also holds an adjunct professorship at Queen’s University, where he earned his Ph.D. in Computer Science from the School of Computing in 2013. He is the founder and director of the IoT Research Lab at Ontario Tech University. Before joining Ontario Tech, he served at the University of Louisiana at Lafayette and the School of Computer Science at Carnegie Mellon University. Dr. Elgazzar was recognized with the Outstanding Achievement in Sponsored Research Award from UL Lafayette in 2017 and the Distinguished Research Award from Queen's University in 2014. He has also received numerous recognitions and best paper awards at leading international conferences.

Dr. Elgazzar is an internationally recognized leader in the fields of Internet of Things (IoT), computer systems, real-time data analytics, and mobile computing. He currently serves as an associate editor for various IEEE/ACM journals and transactions in Peer-to-Peer Networking, Future Internet, IoT, and Mobile Computing. He has chaired several IEEE conferences and symposia on mobile computing, communications, and IoT. A Senior IEEE Member, Dr. Elgazzar actively contributes to technical program and organizing committees for IEEE and ACM events.
\end{IEEEbiography}

\end{document}